\documentclass[aps,twocolumn,prx,floatfix,showpacs,lengthcheck,citeautoscript,superscriptaddress,longbibliography]{revtex4-2}

\usepackage[none]{hyphenat}
\usepackage{amssymb}
\usepackage{amsmath}
\usepackage{graphicx}
\usepackage{bm}
\usepackage{times}
\usepackage{color,soul}
\usepackage[dvipsnames,svgnames,table]{xcolor}
\usepackage{hyperref}
\usepackage{fancyhdr}
\usepackage{enumitem}
\usepackage[english]{babel}
\usepackage[caption=false]{subfig}
\usepackage{braket}
\usepackage{scalerel}

\makeatletter
\let\old@fixname\bbl@fixname
\def\bbl@fixname#1{%
  \@ifundefined{babelalias#1}%
    {\old@fixname{#1}}%
    {\edef\languagename{\csname babelalias#1\endcsname}}}
\makeatother

\hypersetup{
pdfstartview={FitH},
colorlinks=true,    
linkcolor=NavyBlue, 
citecolor=Blue,   
filecolor=NavyBlue, 
urlcolor=NavyBlue   
}

\newcommand{\ve}{\varepsilon}

\newcommand{\bea}{\begin{eqnarray}}
\newcommand{\eea}{\end{eqnarray}}
\newcommand{\Ha}{\hat{\mathcal{H}}}

\newcommand{\ci}{i}

\newcommand{\vf}{v_{\scriptscriptstyle{F}}}
\newcommand{\ef}{\varepsilon_{\scriptscriptstyle{F}}}
\newcommand{\bigdot}{\boldsymbol{\cdot}}
\newcommand{\vre}{\varepsilon}
\newcommand{\vs}[1]{\hat{\bm{#1}}}

\begin{document}
\title{Quantum Hall edge states under periodic driving: a Floquet induced chirality switch}
\author{A. Huam{\'{a}}n}
\affiliation{Centro At{\'{o}}mico Bariloche and Instituto Balseiro,
Comisi\'on Nacional de Energ\'{\i}a At\'omica (CNEA)--Universidad Nacional de Cuyo (UNCUYO), 8400 Bariloche, Argentina}
\affiliation{Instituto de Nanociencia y Nanotecnolog\'{i}a (INN), Consejo Nacional de Investigaciones Cient\'{\i}ficas y T\'ecnicas (CONICET)--CNEA, 8400 Bariloche, Argentina}
\author{L. E. F. {Foa Torres}}
\affiliation{Departamento de F\'{\i}sica, Facultad de Ciencias F\'{\i}sicas y Matem\'aticas, Universidad de Chile, Santiago, Chile}
\author{C. A. Balseiro}
\affiliation{Centro At{\'{o}}mico Bariloche and Instituto Balseiro,
Comisi\'on Nacional de Energ\'{\i}a At\'omica (CNEA)--Universidad Nacional de Cuyo (UNCUYO), 8400 Bariloche, Argentina}
\affiliation{Instituto de Nanociencia y Nanotecnolog\'{i}a (INN), Consejo Nacional de Investigaciones Cient\'{\i}ficas y T\'ecnicas (CONICET)--CNEA, 8400 Bariloche, Argentina}
\author{Gonzalo Usaj}
\affiliation{Centro At{\'{o}}mico Bariloche and Instituto Balseiro,
Comisi\'on Nacional de Energ\'{\i}a At\'omica (CNEA)--Universidad Nacional de Cuyo (UNCUYO), 8400 Bariloche, Argentina}
\affiliation{Instituto de Nanociencia y Nanotecnolog\'{i}a (INN), Consejo Nacional de Investigaciones Cient\'{\i}ficas y T\'ecnicas (CONICET)--CNEA, 8400 Bariloche, Argentina}

\begin{abstract}
We report on the fate of the quantum Hall effect in graphene under strong laser illumination. By using Floquet theory combined with both a low energy description and full tight-binding models, we clarify the selection rules, the quasienergy band structure, as well as their connection with the two-terminal and multi-terminal conductance in a device setup as relevant for experiments. We show that the well-known dynamical gaps that appear in the Floquet spectrum at $\pm\,\hbar\Omega/2$ lead to a switch-off of the quantum Hall edge transport for different edge terminations except for the armchair one, where two terms cancel out exactly. More interestingly, we show that near the Dirac point changing the laser polarization (circular right or circular left) controls the Hall conductance, by allowing to switch it on or off, or even by flipping its sign, thereby reversing the chirality of the edge states. This might lead to new avenues to fully control topologically protected transport.
\end{abstract}
\date{\today}
\maketitle

\section{Introduction}
Forty years ago, the discovery of the precise quantization of the Hall conductance in a two-dimensional electron gas under extreme conditions~\cite{von_klitzing_new_1980} opened the doors to a new chapter in condensed matter physics~\cite{von_klitzing_40_2020}. Elegant topological arguments~\cite{laughlin_quantized_1981,thouless_quantized_1982} explained the precision of the Hall plateaus in practical devices under high perpendicular magnetic fields, while also pointing to new deeper and unifying concepts. Over the last two decades, the use of such topological arguments rapidly expanded~\cite{hasan_colloquium_2010,ando_topological_2013,haldane_nobel_2017} allowing the discovery of, for example, topological insulators in two~\cite{konig_quantum_2007} and three dimensions~\cite{hsieh_topological_2008} and Weyl semimetals~\cite{xu_discovery_2015}. Amid the ever growing family of topological phases, the quantum Hall (QH) effect remains as a paradigmatic case where the topological edge states enjoy the highest degree of robustness, a fact that is nowadays exploited in the new international system of units~\cite{von_klitzing_essay_2019}.

Besides the plethora of manifestations of topological states in (or near) equilibrium conditions, another growing research front aims at using light to change the properties of a material by generating hybrid electron-photon states (also called Floquet-Bloch states) with different spectral and topological properties~\cite{oka_photovoltaic_2009,lindner_floquet_2011,kitagawa_transport_2011,rudner_anomalous_2013,usaj_irradiated_2014,rudner2020,noauthor_quantum_2020}. The latter has become an emerging research front within the so-called quantum materials~\cite{giustino_2020_2020}. Fascinating experiments have unveiled the Floquet-Bloch states~\cite{wang_observation_2013,mahmood_selective_2016} and a much awaited consequence: the light-induced Hall effect~\cite{mciver_light-induced_2020}. While in photonic systems or ultracold matter the experiments allow to reach high driving~\cite{wintersperger_realization_2020} frequencies, which in turn allow suitable theoretical approximations~\cite{eckardt_colloquium:_2017,cooper_topological_2019}, the sweet spot for laser-illuminated Dirac materials corresponds to the (theoretically more challenging) mid-infrared where $\hbar\Omega$ (a few hundreds meV) is much smaller than the bandwidth~\cite{calvo_tuning_2011,perez-piskunow_floquet_2014}. 

Here we address the question of how the QH effect in graphene is affected by laser illumination. Previous studies have mainly focused on the effect of light on the intricacies of the Hofstadter butterfly of different lattices~\cite{kooi_genesis_2018,du_floquet_2018,wackerl_driven_2019,zhao_floquet_2020}, the bulk properties of the irradiated Landau levels and topological invariants~\cite{kooi_genesis_2018}, and related dynamics~\cite{lopez_laser-induced_2015}. By computing the spectrum and the topological invariants, laser-induced modifications on the Hofstadter butterfly and topological properties were recognized. The Hall conductivity in presence of both illumination and an external magnetic field (but without dissipation terms) was also discussed but by means of a generalized Kubo formula~\cite{ding_quantum_2018}. The subject still remains controversial, as the issue of how to properly account for the occupation of the Floquet bands~\cite{dehghani_out--equilibrium_2015,dehghani_dissipative_2014,seetharam_controlled_2015,iadecola_occupation_2015,dehghani_occupation_2016,peralta_gavensky_time-resolved_2018,schuler_how_2020}, specially in this bulk regime when dissipation effects need to be included, has demonstrated to be a difficult task.

In this work we tackle two aspects that are unavoidable in condensed matter experiments: (\textit{i})  the regime of photon energies ($\hbar\Omega$) much smaller than the bandwidth---in particular,  we consider  $\Omega\leq\omega_c$, where $\omega_c$ is the cyclotron frequency; and (\textit{ii}) a multi-terminal device geometry with a laser spot applied to the central part and address the conductance measured in such configurations~\cite{foa_torres_multiterminal_2014}. Specifically, we study the spectrum and the time-averaged conductance both in the two-terminal and multi-terminal case as required for Hall measurements. To such end we use atomistic models within a scattering configuration with a central illuminated spot, thereby allowing for the occupations to be well defined far away in the leads.
The spectrum is first analyzed by using the continuous Dirac model that properly describes the low energy properties of the system. This is done for both zigzag and armchair edge terminations. Such an approach allows us to clearly identify the main features of the Landau-Floquet edge modes. These results are further verified using a more complete tight-binding model which is later used for our transport calculations. The latter are implemented by means of a generalization of the coherent Landauer-B\"uttiker approach to the Floquet picture~\cite{moskalets_floquet_2002,camalet_current_2003,note1}. In this Floquet scattering picture, the leads are not illuminated and have well defined occupations. 

We find that in certain experimentally accessible parameter regions laser illumination leads to important effects including the switch-off of the Hall conductance, the splitting of the Hall plateaus and even a change in the chirality of the propagating states. Interestingly, the Hall conductance is switched off for all edge terminations except for the armchair one, where two contributions cancel out exactly. 

This paper is organized as follows. In Sec.~\ref{Low_energy_hamiltonian} we describe the Landau-Floquet states in graphene within the framework of the Dirac (linear) model.  In Sec.~\ref{tight_binding_model} we study the Landau-Floquet bands in a  tight binding model for zigzag and armchair ribbons. Sec.~\ref{two_terminal_conductance} shows two-terminal conductance simulations which are clarified by visualizing the scattering states. Sec.~\ref{six_terminal_conductance} shows the simulations of the Hall conductance in a  six-terminal configuration. Finally, we summarize our results in Sec.~\ref{conclusions}. 	

\begin{figure}[t]
\includegraphics[width=0.9\columnwidth]{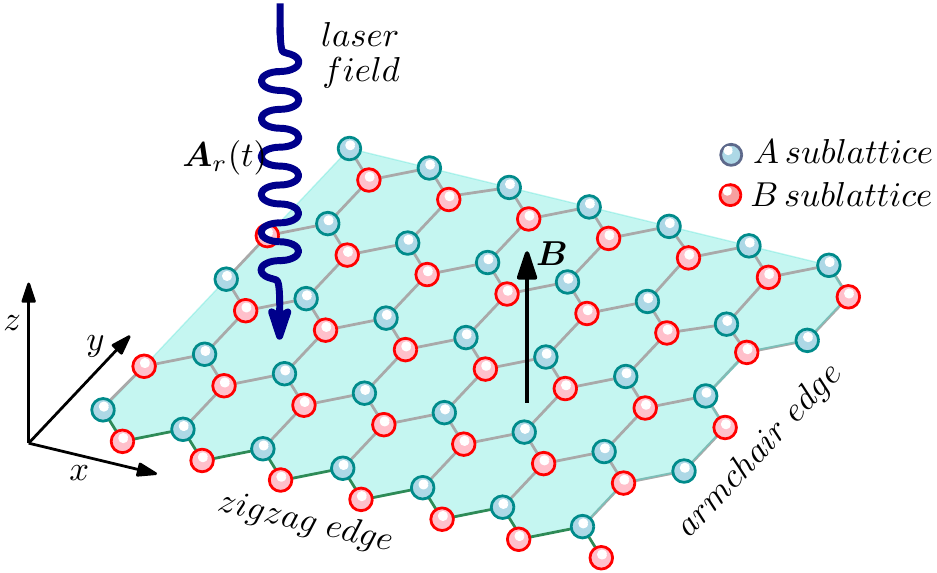}
\caption{(Color online) Geometry used in the Dirac model. The homogeneous magnetic field and the laser are  normal to the graphene monolayer. The zigzag and armchair edges are indicated, highlighting the fact that the former contains only C atoms from a given sublattice ($B$), while the latter contains both.\label{plane}}
\end{figure}
\section{Low energy Hamiltonian\label{Low_energy_hamiltonian}}
The low energy properties of graphene can be described using the following  Hamiltonian~\cite{castro_neto_electronic_2009,Bena2009a},
\begin{equation}
\Ha_0=\vf\,(\tau_z\otimes \sigma_x\,\vs{x}+\tau_0\otimes
\sigma_y\,\vs{y})\bigdot\bm{p}\,,
\label{Dirac_H}
\end{equation}
where $\sigma_i$ ($\tau_i$) with $i=x,y,z$ are Pauli matrices describing the pseudospin (valley) degree of freedom, $\tau_0$ is the $2\times2$ identity matrix, $\bm{p}=p_x\,\vs{x}+p_y\,\vs{y}$ is the momentum operator and $\vf$ is the Fermi velocity.  
The wavefunction $\Psi$ has then four components, $\Psi=[\psi_{AK} , \psi_{BK} , \psi_{AK'} , \psi_{BK'}]^\mathrm{T}$, with amplitudes describing the two inequivalent valleys in the \mbox{Brillouin} zone around $K=(4\pi/3\sqrt{3}a_0,0)$ and $K'=-K$ (the first two amplitudes correspond to $K$ and the remaining ones to $K'$). The parameter $a_0$ is the distance between nearest neighbor carbon atoms.
The presence of a perpendicular magnetic field, $\bm{B}=B\, \vs{z}$, can be described by the well-known Peierls substitution, 
$\bm{p}\rightarrow\bm{p}+\frac{e}{c}\bm{A}$,
with $\bm{A}$ the corresponding vector potential ($-e$ is the electron charge, $e>0$). 
\subsection{The Floquet approach\label{floquet_hamiltonian}}
The illumination with a laser field (applied perpendicularly to the graphene plane) can be modeled as a time-dependent term in the Hamiltonian. Furthermore, as long as the laser is monochromatic (as it will be the case throughout this work), this term is periodic in time and hence it can be treated within the Floquet theory~\cite{grifoni_driven_1998,platero_photon-assisted_2004,kohler_driven_2005}. We briefly describe now this approach before going into its application to our problem.

For a periodic time-dependent Hamiltonian $\Ha(t)$, where $\Ha(t+T)=\Ha(t)$ with the period $T=2\pi/\Omega$, Floquet theory assures the existence of a complete set of solutions of the form $\ket{\Psi_{\alpha}(t)}=\mathrm{e}^{-\ci \varepsilon_{\alpha} t/\hbar}\ket{\phi_\alpha(t)}$ with $\ket{\phi_{\alpha}(t)}=\ket{\phi_{\alpha}(t+T)}$. Replacing this solution in the time-dependent Schr\"odinger equation one obtains:
$\hat{\mathcal{H}}_F \ket{\phi_{\alpha}(t)}=\varepsilon_{\alpha} \ket{\phi_{\alpha}(t)}$, where $\hat{\mathcal{H}}_F=\Ha(t)-i \hbar\,\partial_t$ is called the Floquet Hamiltonian. Thus, we get an eigenvalue equation in the composite space $\mathcal{R} \otimes \mathcal{T}$ (also called Floquet space) where $\mathcal{R}$ is the usual Hilbert space and $\mathcal{T}$ the space of $T-$periodic functions spanned by $\exp{(\ci m \Omega t)}$. The integer $m$ is called the \textit{replica} index. The change in the replica index in a process going from a state with, say, $m$ to  $m+n$ can be assimilated to a number of photon excitations~\cite{shirley_solution_1965}.

Our Hamiltonian can be written as the sum of a time-independent term and a time-dependent one involving the interaction with the laser: $\Ha(t)=\hat{\mathcal{H}}_0+\hat{V}(t)$. By using a Peierls' substitution the time dependent term can be written as
$\hat{V}(t)=\frac{e\,\vf}{c}\,(\tau_z\otimes\sigma_x\,\vs{x}+\tau_0\otimes\sigma_y\,\vs{y})\bigdot\bm{A}_r(t)$, where the vector potential
\begin{equation}\label{laser_A}
\bm{A}_r(t)=A_0\,[\, \cos\alpha\cos\Omega t\,\vs{x}+\sin\alpha\cos(\Omega t-\varphi)\,\vs{y} \,]
\end{equation}
describes the radiation field. For $\varphi=0$, this radiation field is linearly polarized, in which case $\alpha$ is the polarization angle; whereas for  $\varphi=\pi/2$ ($-\pi/2$), and $\alpha=\pi/4$, the radiation is right-handed (left-handed) circularly polarized. The Fourier components of $\hat{V}(t)$, $\hat{V}_{n}=\frac{1}{T}\int_0^T \hat{V}(t) \mathrm{e}^{-\ci n\Omega t} dt$, introduce elements connecting the different Floquet replicas in the Floquet Hamiltonian.

In the low energy approximation there is a further simplification: since  the momentum $\bm{p}$ enters linearly, the perturbation $\hat{V}(t)$ is monochromatic so that its Fourier expansion will have only one harmonic (there is no such a simplification in the tight-binding model, see Sec.~\ref{tight_binding_model}). 
Notice also that any spatial modulation of the laser beam is considered to be larger than all other relevant length scales and hence ignored. This implies that the perturbation cannot mix states that are spatially orthogonal. 
Finally, we write $\hat{V}(t)$ as
\begin{equation}
\label{perturbation_op}
\hat{V}(t)=(\mathcal{V}\,e^{\ci\Omega t} +\mathcal{V}^\dagger\,e^{-\ci\Omega t}),
\end{equation}
with 
\begin{equation}\label{eq2}
\mathcal{V}=\frac{\eta\hbar\Omega}{2}\left[\cos\alpha\,\tau_z\otimes\sigma_x+\sin\alpha\,e^{-\ci\varphi}\,\tau_0\otimes\sigma_y\right]\,.
\end{equation}
Here we have defined the dimensionless parameter $\eta=e\vf A_0/c\hbar\Omega$ that characterizes the intensity of the perturbation. With this, our eigenvalue equation reduces to
\begin{equation}\label{floquetm_reduced}
(\Ha_0+m\hbar\Omega)\ket{\phi_{\alpha,m}}+\mathcal{V}^\dagger|\phi_{\alpha,m+1}\rangle+\mathcal{V}|\phi_{\alpha,m-1}\rangle=\varepsilon_\alpha\ket{\phi_{\alpha,m}}\,,
\end{equation}
where $ \ket{\phi_{\alpha}(t)}=\sum_m e^{\ci m \Omega t} \ket{\phi_{\alpha,m}}$. From this it is clear that the laser field  can only couple replicas $m_i$ and $m_f$ such that $m_f=m_i\pm1$.
For the application of the Floquet formalism, we will expand  $\ket{\phi_{\alpha,m}}$ in a basis of eigenfunctions $\ket{\bm{\chi}_n}$ of the static system (i.e., with the magnetic field alone) 
\begin{equation}\label{expansion_state}
    \ket{\phi_{\alpha,m}}=\sum_n w_{mn}^{(\alpha)}\,|\bm{\chi}_n\rangle\,.
\end{equation}
The eigenstates $|\bm{\chi}_n\rangle$ can be those corresponding to a system with an edge (see Sec.~\ref{zigzag_floquet} and~\ref{armchair_floquet}) or to an infinite (bulk) sample (see Sec.~\ref{bulk_selection_rules}).  From hereon the letter $m$ will be reserved to indicate Floquet replicas. 
\subsection{Zigzag Floquet Hall states \label{zigzag_floquet}}
We start our analysis with the most relevant case of zigzag edges (see Fig.~\ref{plane}). Since in this case the two valleys are not coupled by the boundary condition, we can consider only one of them, say the $K$ valley, and use a simpler  two-component spinor notation. Since this is a generic feature of all terminations except for the armchair one, this case can be considered as the most general. The armchair edge will be analyzed separately later on.
As a basis to expand the Floquet space we use the corresponding QH zigzag edge states, which  are given by (see Appendix \ref{landau_levels_in_graphene} for details)
\begin{equation}\label{eq7}
\bm{\chi}_{\nu_n k}^s(y)=\frac{1}{\sqrt{C_{\nu_n k}}} \left(
 \begin{array}{c}
 D_{\nu_n}(\xi) \\ s \sqrt{\nu_n}D_{\nu_n-1}(\xi)
 \end{array}
 \right),
\end{equation} 
where $\xi{=}\sqrt{2}(y/\ell_B-k\ell_B)$, $\ell_B{=}\sqrt{\hbar c/eB}$ is the magnetic length,  $\varepsilon_n(k)=s\, \hbar\omega_c \sqrt{\nu_n(k)}$ is the energy of the Hall state, where $s=\pm1$ refers to the electron and hole bands, respectively, and $\omega_c{=}\sqrt{2}\vf/\ell_B$ is the cyclotron frequency, $D_\nu(x)$ is the Parabolic Cylinder function of index $\nu$, $k$ is the crystal momentum along the $x$ axis  and $C_{\nu k}$ is a normalization constant.  Here, $n\ge1$ enumerates the positive energy levels, for a given $k$, in ascending order.
Notice that the plane wave factor along the $x$ axis (see Appendix \ref{landau_levels_in_graphene}) can be safely ignored as the perturbation does not mix states with different $k$. 

Since the laser field is monochromatic [cf. Eq.~\eqref{laser_A}], the Floquet matrix $\mathcal{H}_F$---which is a representation of Eq.~\eqref{floquetm_reduced} in the $\mathcal{R} \otimes \mathcal{T}$ space---is an infinite block tridiagonal matrix. With our choice of basis given in  Eq.~\eqref{expansion_state}, the diagonal blocks are itself diagonal. Because we are interested in the effect of the laser field on a few edge states around the Dirac point ($\varepsilon=0$), we will truncate $\mathcal{H}_F$ and retain $2N$ Landau levels, $N$ above and $N$ below the Dirac point and $2M+1$ Floquet replicas. Hence $n=1,\dots,N$ and $m=-M,\dots,M$.
The matrix element between states $| \chi_{\nu_{\tilde{n}} k}^{\tilde{s}}\rangle$  and $|\chi_{\nu_n k}^s \rangle$ in the $m$ and  $m-1$ Floquet replicas, respectively, is simply given  by  
\begin{eqnarray}\label{matrix_element}
\nonumber
\langle \chi_{\nu_{\tilde{n}} k}^{\tilde{s}}|\mathcal{V}|\chi_{\nu_n k}^s \rangle &\!=\!&\frac{\eta \hbar\Omega}{2}\big[
sf(\alpha,\varphi)R_{\nu_n\nu_{\tilde{n}}}{+}
\tilde{s} f(-\alpha,\varphi)R_{\nu_{\tilde{n}}\nu_n}\big]\,,\\
\end{eqnarray}
where  $f(\alpha,\varphi)=\cos\alpha-\ci\sin\alpha\,e^{-\ci\varphi}$ and 
\begin{equation}
R_{\nu\nu'}=\frac{\ell_B\sqrt{\nu}}{\sqrt{2\,C_{\nu k}C_{\nu' k}}}\int_{-\sqrt{2}k\ell_B}^\infty\!\mathrm{d}\xi\, D_{\nu-1}(\xi)D_{\nu'}(\xi)\,.
\end{equation}
Similar calculations can be performed in the $K'$ valley using the eigenfunctions given by Eq.~\eqref{basisKp}.
\begin{figure}[t]
\includegraphics[width=0.95\columnwidth]{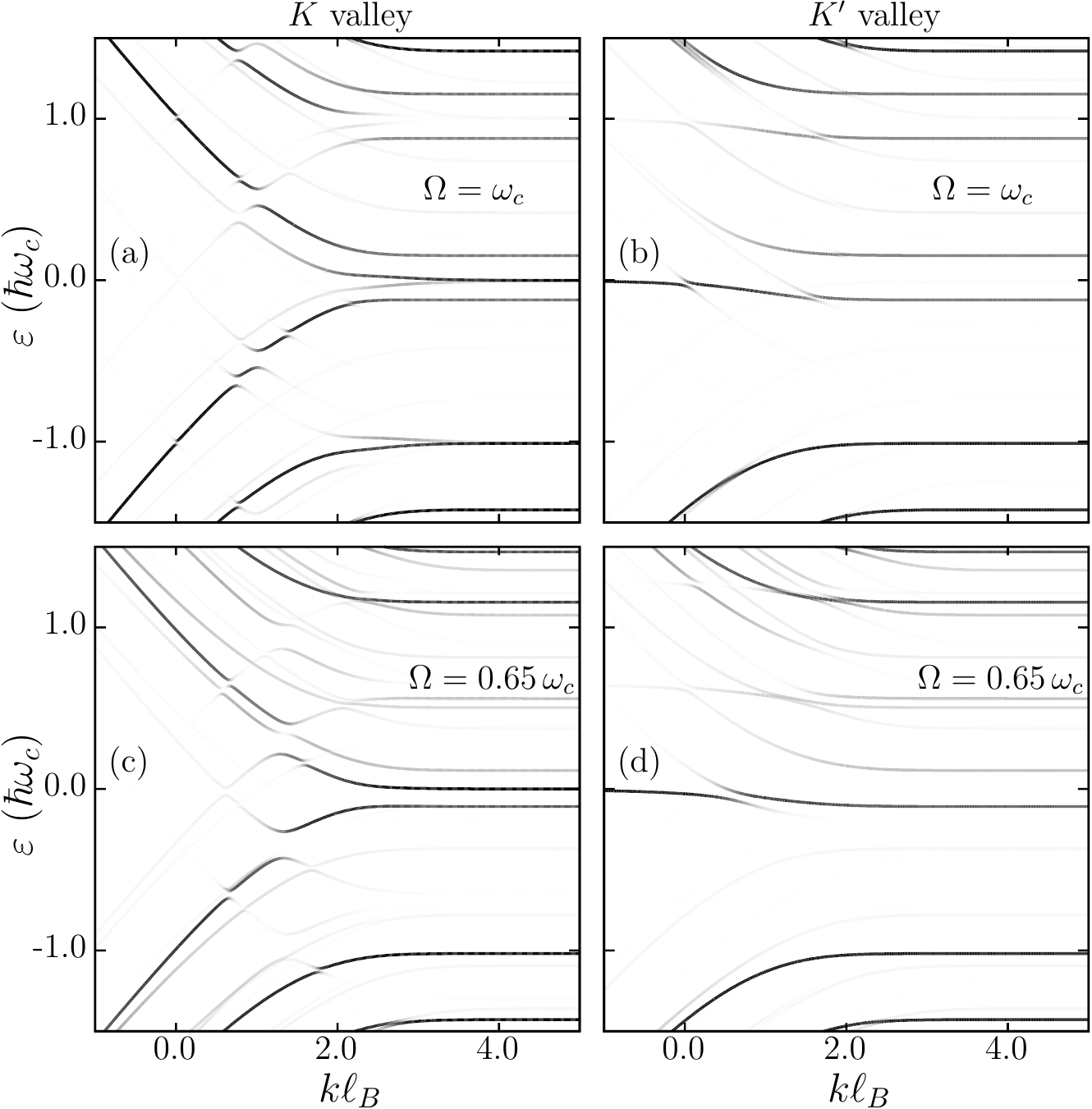}
\caption{Quasienergy spectrum projected on the $m=0$ Floquet replica [$\rho_0(\varepsilon,k)$, solid black lines] as a function of the dimensionless wavevector $k\ell_B$ along a zigzag edge irradiated with a circularly polarized laser. 
For (a) and (b) [(c) and (d)]  we use $\eta=0.2\,\omega_c/\Omega$ ($\eta=0.3\,\omega_c/\Omega$).  Subplots (a) and (b)    correspond to the resonant case ($\Omega=\omega_c$) for valleys $K$ and $K'$, respectively. Similarly, (c) an (d) correspond to a non resonant photon energy ($\Omega=0.65\omega_c$).  Here five Floquet channels ($-2\leq m\leq 2$) were used.  \label{zigzag_Floquet_spectrum}}
\end{figure}

Let us now consider the case of a laser field with positive (counter clockwise) circular polarization: $\alpha=\pi/4$ and $\varphi=\pi/2$. Then we have $f(\pi/4,\pi/2)=0$ and $f(-\pi/4,\pi/2)=\sqrt{2}$, and thus the right hand side of Eq.~\eqref{matrix_element} reduces to $\eta\hbar\Omega\,\tilde{s}R_{\nu_{\tilde{n}}\nu_n}/\sqrt{2}$. Figure~\ref{zigzag_Floquet_spectrum} shows the quasienergy dispersion of the Floquet Hall edge states, weighted by their projection on the $m=0$ Floquet replica. 
These dispersion relations were obtained by numerically  calculating the following spectral density
\begin{equation}
\label{rho}
    \rho_0(\varepsilon,k)=-\frac{1}{\pi}\mathrm{Im}\,\mathrm{Tr}_0\left[ \varepsilon+\ci0^+-\mathcal{H}_F(k)\right]^{-1}\,,
\end{equation}
where the trace $\mathrm{Tr}_0$ is taken only over the $m=0$ subspace. Here we used five Floquet replicas  ($M=2$), twelve Landau levels ($N=6$) and two different photon energies: (i) resonant with the first bulk Landau level ($\Omega=\omega_c$) and (ii) off-resonant ($\Omega=0.65\,\omega_c$). We include also the results for the $K'$ valley which were obtained in a similar fashion.

The main new features in the spectrum that are apparent from the Fig.~\ref{zigzag_Floquet_spectrum} are: (i) the splitting of the bulk Landau levels and the lack of electron-hole symmetry, both analyzed in detail in Sec.~\ref{bulk_selection_rules}; (ii) the appearance of multiple dynamical gaps (or, more precisely, avoided crossings) of different order in $\eta$.  In particular, the first order ones at $\pm\,\hbar\Omega/2$ in the $K$ valley arises from the resonant coupling between the lowest electron and the highest hole edge states and, as we will show when discussing the transport properties, lead to the (almost complete) suppression of the  QH conductance; (iii) the bending of the otherwise flat zero energy state ($\nu=0$) of the $K'$ valley, which results from the direct coupling to the lowest Landau level of the electron band $\ket{\bm{\chi}_{\nu_1k}^+}$ (in the Floquet picture it corresponds to coupling to the  $m=-1$ replica). This leads to an edge mode with a polarization dependent dispersion that it is \textit{always} a counter-propagating mode, in the sense that it has the opposite velocity that the edge states with the same sign of quasienergy. This, in turn, causes a change of the sign of the Hall conductance, as we discuss in Section \ref{six_terminal_conductance}; (iv) in a small quasienergy region above $\ve=\hbar\omega_c$ there is an effective reduction of the number of edge states as the one coming from the $K'$ valley is shifted upwards. In a finite sample, the same happens for the $K$ valley on the other edge. This leads to the emergence of a $4e^2/h$ feature in the two terminal conductance as discussed in Section \ref{two_terminal_conductance}.

The size $\Delta_\mathrm{dym}$ of a dynamical gap is given, to first order in $\eta$, by the matrix element between the two states involved in the avoided crossing [Eq.~\eqref{matrix_element}]. In the case of the gap at $\varepsilon=\hbar\Omega/2$, they are $\ket{\bm{\chi}_{\nu_1k}^+}$ and $\ket{\bm{\chi}_{\nu_1k}^-}$ in the $m=0$ and $m=1$ replicas, respectively, so that
\begin{equation}\label{dynamical_gap}
\Delta_\mathrm{dym}\simeq\eta\,\hbar\Omega\,\sin\alpha|R_{\nu_1\nu_1}|\,,
\end{equation}
where $\nu_1(k)=(\Omega/2\omega_c)^2$ defines the value of $k$ where the resonant condition is fulfilled. 
Eq.~\eqref{dynamical_gap} is not restricted to a circularly polarize laser. In fact, it must be  noted that in the case of a linearly polarized beam,  $\Delta_\mathrm{dym}$ depends on the relative orientation of the  electric field  and the edge, being zero if the electric field is parallel to the edge ($\alpha{=}0$).
\subsection{Armchair Floquet Hall states\label{armchair_floquet}}
Now we apply a similar treatment to the case of armchair edges. For this special termination, however, one needs to take into account both valleys at the same time~\cite{castro_neto_electronic_2009}, as the boundary condition mixes them. The eigenfunctions for the static system are presented in Appendix \ref{landau_levels_in_graphene}. They are now four-component spinors
\begin{equation}\label{arm_eq}
\bm{\chi}_{\nu_n k}^s(x)=\frac{1}{\sqrt{C_{\nu_n k}}}
\left[\begin{array}{c}
-\ci s \tau_n D_{\nu_n}(\xi)e^{\ci Kx} \\
-\tau_n\sqrt{\nu_n} D_{\nu_n-1}(\xi)e^{\ci Kx} \\
\ci s\sqrt{\nu_n} D_{\nu_n-1}(\xi)e^{-\ci Kx} \\
 D_{\nu_n}(\xi)e^{-\ci Kx}
\end{array}\right],
\end{equation}
with $\xi=\sqrt{2}(x/\ell_B-k\ell_B)$. The eigenenergies are given by $\varepsilon_n=s\,\hbar\omega_c\sqrt{\nu_n}$ with $s=\pm 1$, the meaning of $s$ being the same as in the zigzag case. The new parameter $\tau_n=(-1)^{n+1}$ indicates the {\it branch} $\nu_n$ belongs to [see Appendix~\ref{landau_levels_in_graphene} for details as well as for the explicit form of $C_{\nu_n k}$].  Using Eq.~\eqref{arm_eq} and the interaction matrix Eq.~\eqref{eq2} we have the following matrix element that enters in $\mathcal{H}_F$
\begin{eqnarray}\label{armchair_selection_rule_general}
\bra{\bm{\chi}^{\tilde{s}}_{\nu_{\tilde{n}}k}}\mathcal{V}\ket{\bm{\chi}^s_{\nu_nk}}&=&\ci\frac{\eta\hbar\Omega}{2}(\tilde{s}s+\tau_{\tilde{n}}\tau_n)\times \\
\nonumber
&&[sf(-\alpha,\phi)R_{\nu_{\tilde n} \nu_n} - \tilde{s}f(\alpha,\phi)R_{\nu_n \nu_{\tilde n} }]\,.
\end{eqnarray}
This gives an interesting selection rule: $\bra{\bm{\chi}^{\tilde{s}}_{\nu_{\tilde{n}}k}}\mathcal{V}\ket{\bm{\chi}^s_{\nu_nk}}=0$ if $\tilde{s}s=-\tau_{\tilde{n}}\tau_n$. In particular, for the $n=\tilde{n}=1$ edge mode the coupling between the conduction  ($c$, $s=1$) and valence ($v$, $\tilde{s}=-1$) bands vanishes
\begin{equation}\label{armchair_selection_rule}
\bra{\bm{\chi}^v_{\nu_{1}k}}\mathcal{V}\ket{\bm{\chi}^c_{\nu_1 k}}=0\,.
\end{equation}
Because this matrix element is responsible for the opening of a dynamical gap  at $\pm\,\hbar\Omega/2$, we do not expect such a gap in the armchair case---of course, this argument refers to a first order gap; higher order (smaller) gaps in fact exist at these points. Note that, in general, there is no coupling between electron and hole levels (i.e. $s\tilde{s}=-1$) belonging to the same solution branch (i.e. $\tau_{\tilde{n}}\tau_n=1$). Similarly, $\langle \bm{\chi}^v_{\nu_{\tilde{n}} k} |\mathcal{V}| \bm{\chi}^v_{\nu_n k}\rangle=\langle \bm{\chi}^c_{\nu_{\tilde{n}} k} |\mathcal{V}| \bm{\chi}^c_{\nu_n k}\rangle=0$   if $\tau_n=-\tau_{\tilde{n}}$. i.e., same electron or hole  character and different solution branch. All this implies that, considering only first order couplings,  the armchair edges have more symmetries than the zigzag ones, leading to a simpler Floquet spectrum. 
\begin{figure}[t]
    \centering
    \includegraphics[width=0.95\columnwidth]{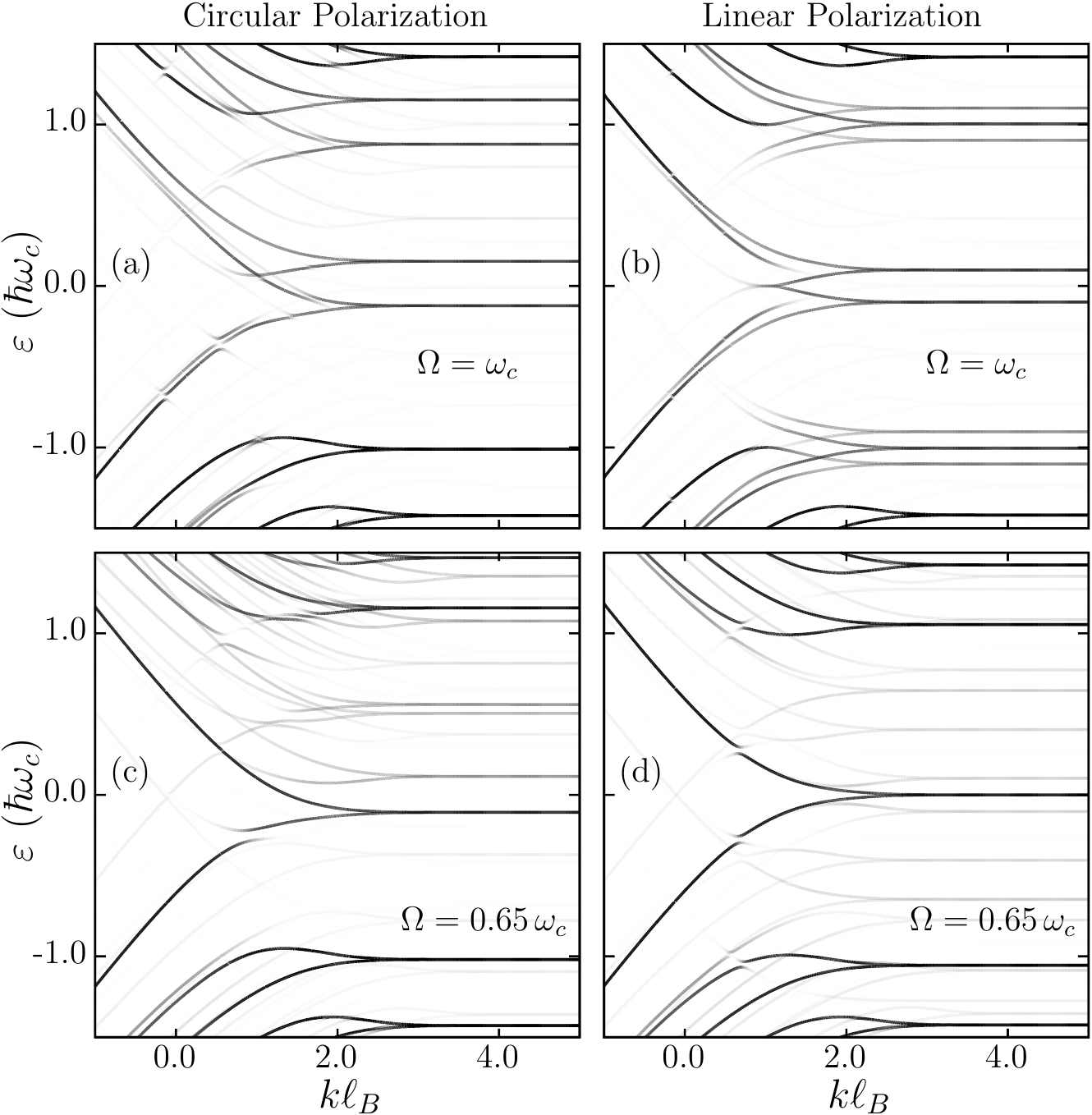}
    \caption{Quasienergy spectrum projected onto the replica $m=0$ for an irradiated armchair edge. The laser is circularly (linearly) polarized  in (a)  and (c) [(b) and (d)]. 
    In (a) and (b) the laser is in resonance ($\Omega=\omega_c$), with $\eta=0.2\,\omega_c/\Omega$; whereas in (c) and (d) it is out of resonance ($\Omega=0.65\omega_c$), with $\eta=0.3\,\omega_c/\Omega$. Five Floquet replicas ($-2\leq m \leq 2$) were  included. It is clear, in comparison with Fig.~\ref{zigzag_Floquet_spectrum}, the absence of a first order dynamical gap at $\pm\,\hbar\Omega/2$. Higher order gaps due to the coupling with $m=\pm2$ replicas are present. 
   }
    \label{armchair_dirac}
\end{figure}

The corresponding weighted Floquet spectrum, calculated with the projected spectral density $\rho_0(\varepsilon,k)$, is shown in Fig.~\ref{armchair_dirac} for a circular ($\varphi=2\alpha=\pi/2$) and linear ($\alpha=\varphi=0$) polarization of the laser field. As in the previous section we use five Floquet replicas ($M=2$). The most striking difference with Fig.~\ref{zigzag_Floquet_spectrum} is the lack of first order  dynamical gaps at $\pm\,\hbar\Omega/2$, in complete agreement with Eq.~\eqref{armchair_selection_rule}. Moreover, when the photon energy is out of resonance [Figs.~\ref{armchair_dirac}(c) and \ref{armchair_dirac}(d)], the weighted Floquet bands are similar to the ones of the static system, except for an energy shift of certain Landau levels. It is interesting that certain gaps appear at the crossing of the static system ($m=0$) with replicas $m=\pm 2$. Since our linear model only couples directly Floquet channels differing in one photon, these gaps are of second order and thus smaller than those seen in Fig.~\ref{zigzag_Floquet_spectrum} at $\vre=\pm\,\hbar\Omega/2$. The flat states for $k\ell_B\gg2$ correspond to the bulk Landau levels, so that their shifting and splitting follows the pattern of the latter, which we now discuss.
\subsection{Bulk selection rules \label{bulk_selection_rules}}
To better understand some of the features observed in Figs. \ref{zigzag_Floquet_spectrum} and \ref{armchair_dirac},  it is useful to analyze the bulk case.
For that we calculate $\rho_0(\varepsilon)$ [cf. Eq.~\eqref{rho}] with $\mathcal{H}_F$ written in the basis of the bulk eigenfunctions (see Appendix \ref{landau_levels_in_graphene}).
The matrix elements of $\mathcal{H}_F$ between the Floquet bulk eigenfunctions can be calculated by considering each valley separately as they are decoupled in bulk. Using the solutions $|\bm{\chi}_{lk}\rangle$ given in Eq.~\eqref{bulk_function} for the $K$ valley (and omitting the superscript $K$) we obtain the following matrix elements
\begin{align}\label{selection_rules}
&\langle  \bm{\chi}_{nk}|\mathcal{V}|\bm{\chi}_{lk}\rangle=\frac{\eta\hbar\Omega}{4\sqrt{1-(\delta_{n0}+\delta_{l0})/2}}\times\,\notag\\
&\Big[ f(\alpha,\varphi)\mbox{sgn}(l)\delta_{|n|,|l|-1}+ f(-\alpha,\varphi)\mbox{sgn}(n)\delta_{|n|,|l|+1} \Big]\,.
\end{align}
In the right-handed circularly polarized case, $f(\alpha,\varphi)=0$ and this matrix element is proportional to $\delta_{|n|,|l|+1}$. In this way we obtain the selection rule  $|l|=|n|-1$, where $l$ is the Landau index of the state with an extra absorbed photon [$(l,m+1)\leftrightarrow (n,m)$ transition in the Floquet space]. For the opposite circular polarization the $|l|=|n|+1$ rule applies.
When the polarization is linear, Eq.~\eqref{selection_rules} dictates that $||n|-|l||=1$. The matrix elements between eigenfunctions in the $K'$ valley are the same as those in Eq.~\eqref{selection_rules}, and thus the same selections rules apply.
\begin{figure}[t]
\includegraphics[width=0.95\columnwidth]{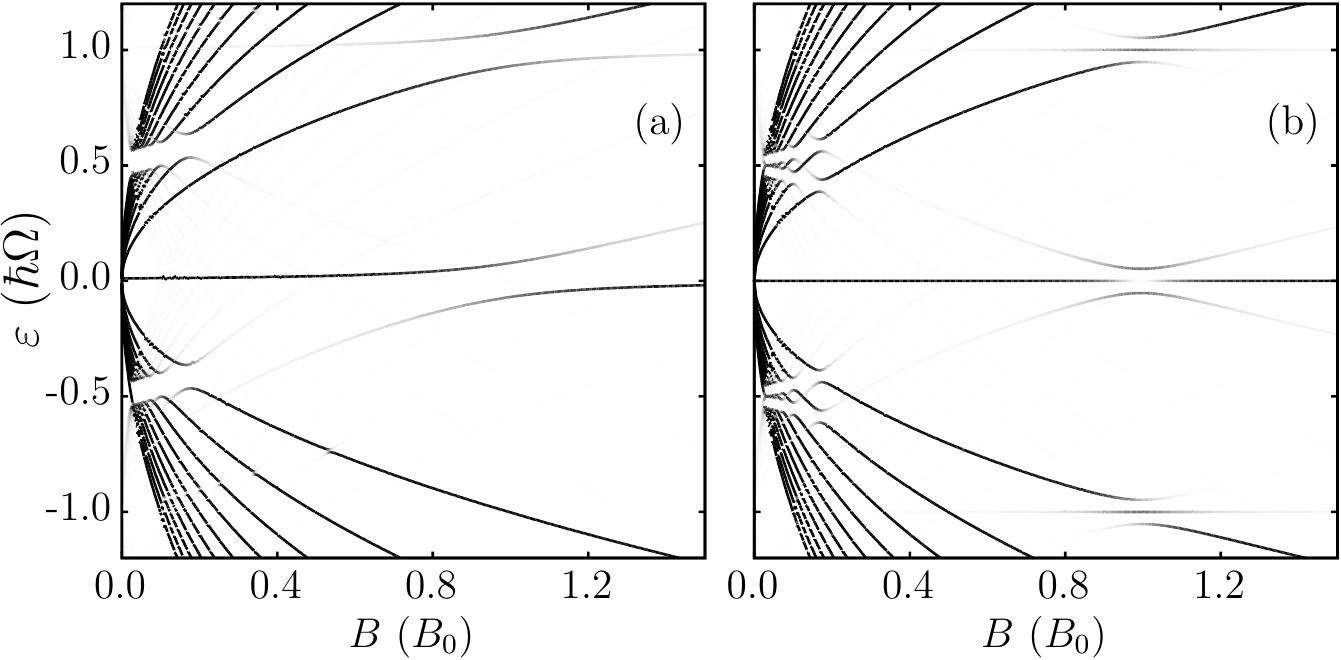}
\caption{Weighted bulk Landau-Floquet spectral density $\rho_0(\varepsilon,B)$ as a function of the  magnetic field for a fixed value of $\Omega$. In (a) and (b) the laser field is circularly ($\alpha=\pi/4$, $\phi=\pi/2$) and linearly ($\alpha=\pi/2$, $\phi=0$) polarized, respectively. Here $\eta=0.15$ and $B_0=\Omega^2\hbar c/(2e\vf^2)$.\label{bulk_gaps}} 
\end{figure}

These selection rules can be clearly seen when we plot the Floquet spectral density  $\rho_0(\vre)$ as a function of $B$ for a fixed value of $\hbar\Omega$, as shown in Fig.~\ref{bulk_gaps}.
The calculations were carried out with twenty Landau levels ($N=10$,), five Floquet replicas ($M=2$) and  $\eta=0.15$. To properly scale the spectrum  it is useful to define an auxiliary  magnetic field $B_0=\hbar c \Omega^2/(2e\vf^2)$ so that $\omega_c/\Omega=\sqrt{B/B_0}$  and the energy of the $m$-th Floquet replica is simply $\varepsilon_n^{(m)}=\left(\mbox{sgn}(n)\sqrt{\frac{B}{B_0}|n|}+m\right)\hbar\Omega$.
Figure~\ref{bulk_gaps}(a) shows the case of a circular polarization. First we notice that, as $B$ approaches zero, there are dynamical gaps of almost constant size that tend to center around $\varepsilon=\pm\,\hbar\Omega/2$. 
These are reminiscences of the well-known dynamical gaps of irradiated graphene that appear in the absence of any magnetic field (with size  roughly equal to $\eta\hbar\Omega$) \cite{Oka2009,calvo_tuning_2011}. Another interesting feature is the appearance of anti-crossings near $B=B_0$, that is, when the laser field is in resonance with the transition between the zero and the first Landau level. In the case of the anti-crossing near $\varepsilon=0$, it originates from the degeneracy of the Floquet states $|\bm{\chi}_{0k},0\rangle$ and $|\bm{\chi}_{1k},-1\rangle$, as dictated  by the selection rules---here we use the notation $\ket{\bm{\chi}_{\nu k},m}$ to indicate the $m-$th Floquet replica the state $\ket{\bm{\chi}_{\nu k}}$ belongs to---while for the one at $\varepsilon=\hbar\Omega$ it corresponds to the near degeneracy between $|\bm{\chi}_{1k},0\rangle$ and $|\bm{\chi}_{0k},1\rangle$. Note that because of the selection rules there is no coupling between $|\bm{\chi}_{-1k},0\rangle$ and $|\bm{\chi}_{0k},-1\rangle$ and so $\rho_0(\vre)\neq\rho_0(-\vre)$ (there is no electron-hole symmetry). This is consistent with the results of Refs.~\cite{zhao_floquet_2020,wackerl_driven_2019} where the full Hofstadter butterfly spectrum (tight-binding model) was analyzed. 
Finally, we mention that the series of gaps near $\hbar\Omega/2$  arises from the anti-crossings between the Floquet states $|\bm{\chi}_{nk},0\rangle$ and  $|\bm{\chi}_{1-n,k},1\rangle$ with $n=1,2,\cdots$, while those  near $-\hbar\Omega/2$ appear at the crossings of $|\bm{\chi}_{-nk},0\rangle$ and  $|\bm{\chi}_{n+1,k},-1\rangle$.

As we have already mentioned, for a linearly polarized laser the selection rules require $||n|-|l||=1$, with both $n$ and $l$ entering symmetrically. This implies that $\rho_0(\vre)=\rho_0(-\vre)$  as it is clear from Fig.~\ref{bulk_gaps}(b). It is interesting to analyse  in particular the triple crossing that occurs near $\vre=0$ for $B=B_0$ (resonance condition, $\Omega=\omega_c$). In that case,  $|\bm{\chi}_{0k},0\rangle$, $|\bm{\chi}_{-1k},1\rangle$ and $|\bm{\chi}_{1k},-1\rangle$ become degenerate while the selection rules allow the coupling between $|\bm{\chi}_{0k},0\rangle$ and each of the other two states, with a matrix element $\eta\hbar\Omega\,e^{\pm\ci\alpha}/2\sqrt{2}$, respectively. Within this restricted subspace, a straightforward diagonalization gives the eigenvalues $\lambda_0=0$ and $\lambda_\pm=\pm\,\eta\hbar\Omega/2$. The eigenvector corresponding to $\lambda_0$ is $(|\bm{\chi}_{-1k},1\rangle+|\bm{\chi}_{1k},-1\rangle)/\sqrt{2}$ which does not have any weight on the $m=0$ replica, as it is evident from the lack of spectral weight shown in Fig.~\ref{bulk_gaps}(b). For the other eigenvalues $\lambda_\pm$ we have that the corresponding eigenvectors are $(\sqrt{2}|\bm{\chi}_{0k},0\rangle \mp e^{\ci\alpha}(|\bm{\chi}_{-1k},1\rangle+|\bm{\chi}_{1k},-1\rangle)/2$, both with the same weight ($1/2$) on the $m=0$ replica. A similar calculation explains the features observed at $\vre=\pm\,\hbar\Omega$ in Fig.~\ref{bulk_gaps}(b).
Finally, the small gaps near $\hbar\Omega/2$ that appear at low $B$ result from the crossing of $|\bm{\chi}_{nk},0\rangle$ with $|\bm{\chi}_{-(n+1)k},1\rangle$ and $|\bm{\chi}_{-(n-1)k},1\rangle$  ($n=1,2,\cdots$), and similarly for the gaps near $-\hbar\Omega/2$. It is worth mentioning here that in the absence of a  magnetic field there is no laser induced dynamical gap in $\rho_0(\vre)$ for a linearly polarized laser but a pseudo-gap that closes linearly in energy at exactly $\vre=\hbar\Omega/2$~\cite{calvo_tuning_2011}. This can be seen in Fig.~\ref{bulk_gaps}(b) as $B$ goes to zero, where the size of such gaps become smaller until they vanish at $B=0$. In this case, the presence of this pseudo-gap is reveled by the Landau-Floquet states that appear nearly pinned at $\vre=\hbar\Omega/2$.
\section{Tight binding model \label{tight_binding_model}}
\begin{figure}[t]
\begin{center}
\includegraphics[width=0.9\columnwidth]{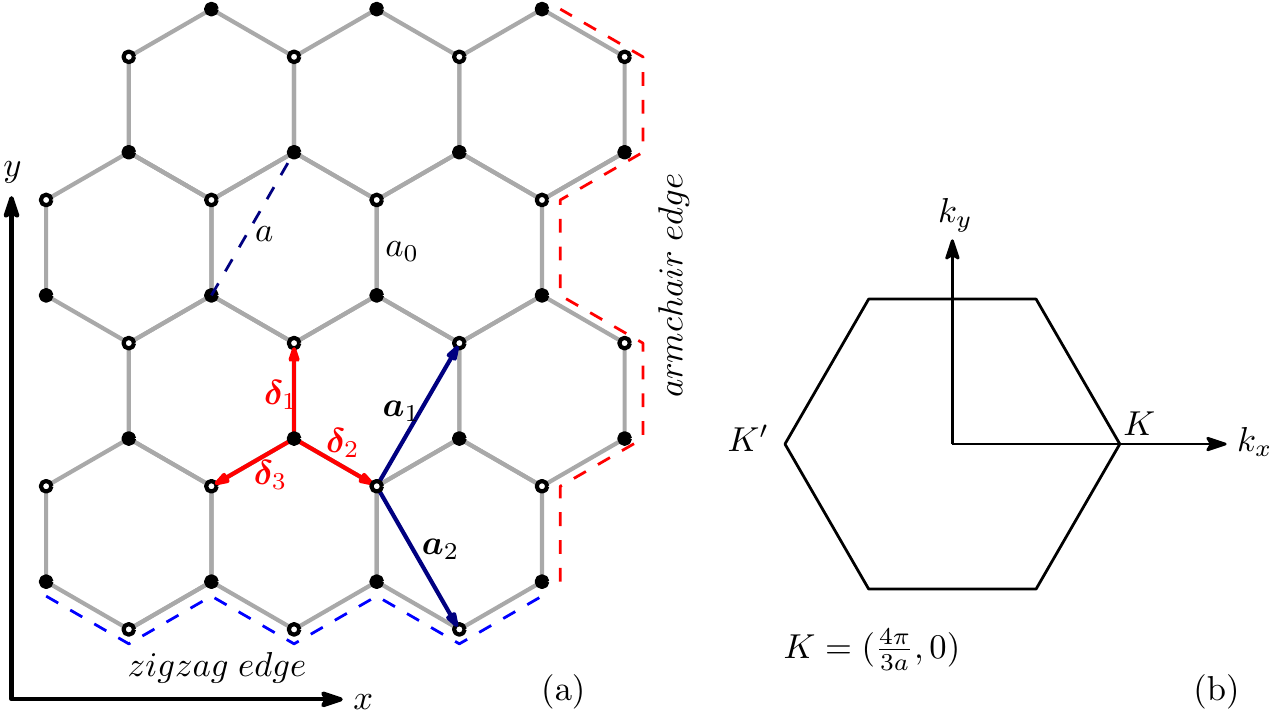}
\caption{(Color online) (a) Geometry of the problem in the tight-binding model, with the lattice $\bm{a}_j$ and nearest neighbors $\bm{\delta}_j$ vectors indicated, whose modules are $a$ and $a_0$, respectively ($a=\sqrt{3}\,a_0$). We refer all our calculations to this configuration, and thus a zigzag (armchair) ribbon has a translational symmetry along the $x$ ($y$) direction. (b) First Brillouin Zone and the two non equivalent valleys $K$ and $K'$.\label{lattice2}}
\end{center}
\end{figure}

The Dirac model is suitable only for describing the low energy excitations near the Dirac point ($\varepsilon=0$), where the energy dispersion is almost conical. A better and more complete description is given by a tight-binding (TB) model, where the $p_z$ carbon orbitals in graphene are described by $\hat{\mathcal{H}}=\sum_{\langle i,j \rangle} t_{ij}\,\hat{c}_{i}^\dagger \hat{c}_{j}+h.c.$. Here $\hat{c}_{j}$ is a destruction operator at the position $j$, the notation $\langle i,j \rangle$ implies that the summation is carried over nearest neighbors only, separated by a distance $a_0=1.42$\AA\, (see Fig.~\ref{lattice2}), while the hopping $t_{ij}$ is independent of the site: $t_{ij} =t=-2.8\,$eV.
The effect of an external field described by the vector potential $\bm{A}(\bm{r},t)$ is included as before via the Peierls substitution, which in the TB approach is given by
\begin{equation}\label{peierls}
t_{ij} \rightarrow t_{ij}\times \exp{\left[\left(\frac{\ci e}{\hbar c}\right)\int_{\bm{r}_i}^{\bm{r}_j} \bm{A}(\bm{r},t)\cdot d\bm{r}\right]}\,.
\end{equation}

\subsection{Landau levels}
Let us first briefly describe the well-known effects of an homogeneous magnetic field  on a graphene ribbon with either zigzag or armchair edges. Figure~\ref{lattice2} shows that with our choice of axes the ribbon has  translation symmetry along the $x$ ($y$) direction for a zigzag (armchair) edge. This symmetry allows us to introduce a Bloch function with a crystal momentum $k_x$ or $k_y$ along the relevant symmetry direction. In order to preserve this symmetry we choose the gauge $\bm{A}(x)=-By\,\vs{x}$ [$\bm{A}(x)=Bx\,\vs{y}$] for the zigzag (armchair) ribbon.
Hence, the Peierls substitution for the zigzag ribbon is explicitly given by  $t_{ij} = t\,\exp[-\ci\,\zeta(x_j-x_i)(y_j+y_i)/a_0^2]$, where  $\zeta=\pi\Phi/\Phi_0$, $\Phi=Ba_0^2$ and $\Phi_0=hc/e$ is the flux quantum.

\begin{figure}[t]
\includegraphics[width=0.9\columnwidth]{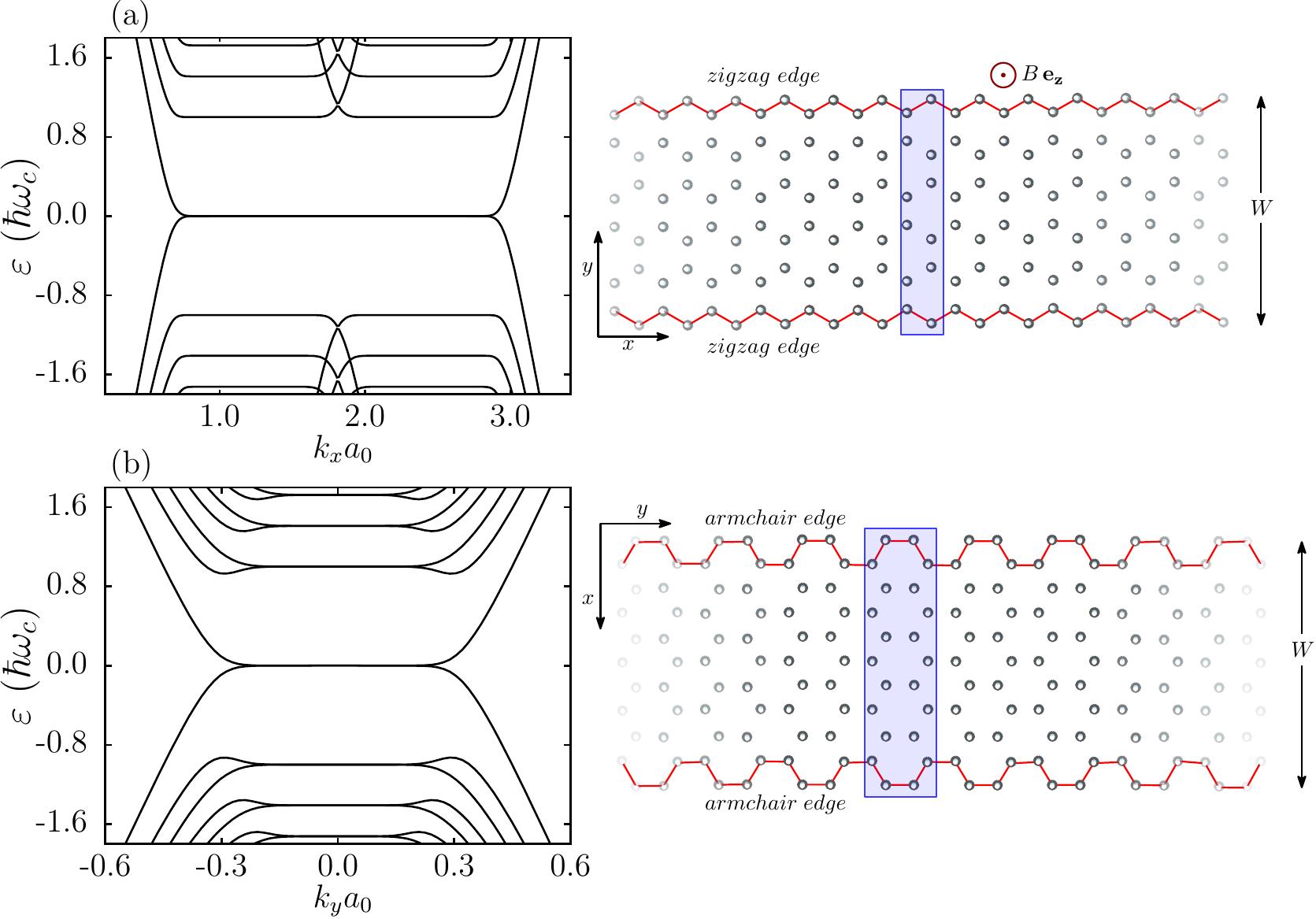}
\caption{(Color online) Left panels: Energy dispersion of the Landau  levels for  (a) a zigzag  and (b) an armchair ribbon with widths $W=224\,a_0$ and $W=75\sqrt{3}\,a_0$, respectively, and $\zeta=0.003$ (equivalent to $\hbar\omega_c\simeq0.46\,$eV and $\ell_B\simeq 13\,a_0$).  Right panels: Geometry of the ribbons with the corresponding unit cells highlighted by the purple box. The magnetic field $B\vs{z}$ is applied normally to the ribbon's plane. 
\label{ribbon}}
\end{figure}

Figure \ref{ribbon}(a)  shows the Landau  bands for a zigzag ribbon of a width of $300$ atoms ($W=224a_0$) and $\zeta=0.003$. There are bulk Landau levels (flat bands), as well as dispersive edge states due to the confinement imposed by the ribbon. The dispersion relations in each Dirac point is in very good agreement with those found with the Dirac model [compare with Fig.~\ref{zigzag_dispersion}(a) and~\ref{zigzag_dispersion}(b) in Appendix~\ref{landau_levels_in_graphene}]. It is worth mentioning here that not all the zero energy states in this geometry are bulk Landau levels. There is also a trivial dispersionless edge mode that appear on zigzag ribbons in the absence of a magnetic field (see Appendix~\ref{landau_levels_in_graphene} for a further discussion). These modes, being dispersionless, are not affected by the Lorentz force. 

A similar calculation done for the armchair case leads to the Landau spectrum shown in Fig.~\ref{ribbon}(b). Here the ribbon is $302$ atoms wide ($W=75\sqrt{3}\,a_0$). 
In contrast with a zigzag ribbon, the two dispersionless states at $\vre=0$ are fully located in bulk, and are identified with the $n=0$ Landau level.

\subsection{Floquet states}
We now add a time dependent laser field using the vector potential given in Eq.~\eqref{laser_A}, which  is assumed to be  homogeneous throughout space.  The integral in Eq.~\eqref{peierls} is then simply $\bm{R}_{ij}\cdot \bm{A}(t)$, where $\bm{R}_{ij}=a_0\,(\cos\theta_{ij}\,\vs{x}+\sin\theta_{ij}\,\vs{y})$ is the vector connecting neighboring sites $i$ and $j$. For a laser field with positive circular  polarization, the Peierls substitution leads to $t_{ij}= t\, \exp[\ci z \cos(\Omega t-\theta_{ij})]$, while for the linearly polarized case we have $t_{ij}= t\, \exp[\ci z \cos(\theta_{ij}-\alpha) \cos\Omega t]$. Here we have introduced the dimensionless quantity $z=ea_0A_0/\hbar c=(\Omega a_0/\vf)\,\eta$ that measures the intensity of the perturbation. It is clear then that the time-dependent TB Hamiltonian for the irradiated ribbon is periodic in time but not  harmonic. In the Floquet formulation, the Floquet matrix elements now couple replicas with $\Delta m\neq\pm1$. The Fourier components of the hamiltonian can be calculated using the well-known Jacobi-Anger identity: $e^{\ci r\cos\theta}=\sum_{m=-\infty}^{+\infty} \ci^m J_m(r)\,e^{im\theta}$, 
where $J_m(r)$ are the Bessel functions of the first kind of integer order. $\mathcal{H}_F$, which is  no longer block tridiagonal, is truncated to a finite number of Floquet channels for its numerical diagonalization. In the following we retain five Floquet replicas $-2\leq m\leq 2$ (unless otherwise stated) and calculate  $\rho_0(\varepsilon,k)$ by means of Eq.~\eqref{rho}. This number of replicas guarantees that, for the value of the parameters we use, the most relevant features in $\rho_0(\varepsilon,k)$ are well described.

\begin{figure}[t]
\includegraphics[width=0.95\columnwidth]{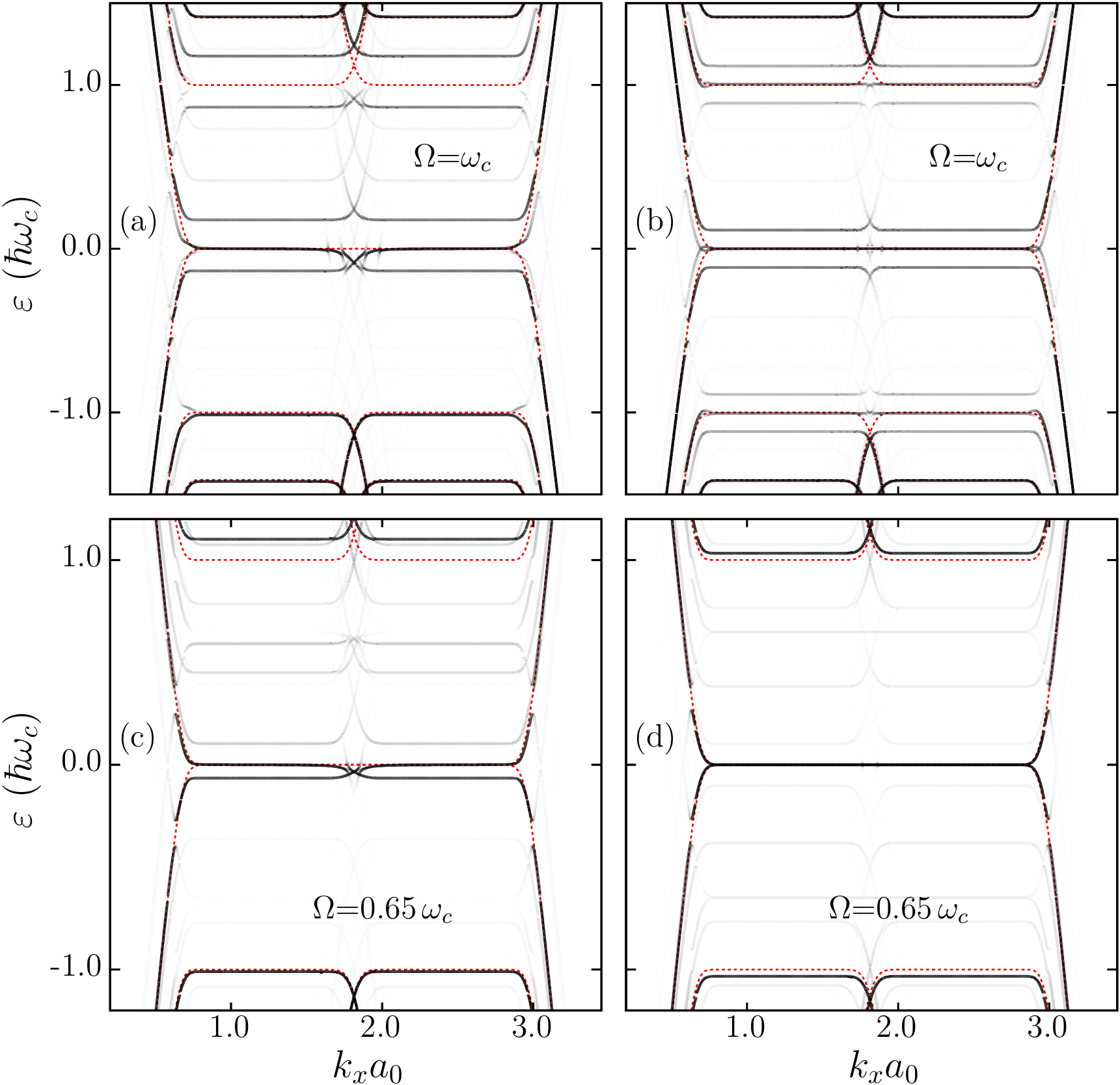}
\caption{(Color online) Landau-Floquet spectral density $\rho_0(\vre,k)$ in the presence of an homogeneous magnetic field ($\zeta=0.003$) and a laser with intensity $z=0.025$. The polarization is circular ($\varphi=2\alpha=\pi/2$) for (a) and (c), and linear ($\alpha=\pi/2,\;\varphi=0$) for (b) and (d).   Sub-plots (a) and (b) correspond to the resonant case with $\Omega=\omega_c$, whereas in (c) and (d) the laser field is off-resonant with $\Omega=0.65\,\omega_c$. Five Floquet replicas ($-2\leq m\leq 2$) were used \label{zigzag_ribbon_Floquet}. The ribbon is $300$ atoms wide ($W=224\,a_0$).}
\end{figure}
The Landau-Floquet spectral density for a zigzag ribbon of width $W=224\,a_0$ is shown in Fig.~\ref{zigzag_ribbon_Floquet}: (a) and (c) correspond to  the circularly polarized case, while (b) and (d)  to a linearly polarized laser, with $\alpha=\pi/2$---the direction of polarization is perpendicular to the edges of the ribbon---which guarantees a maximum size in the gap opening, see Eq.~\eqref{dynamical_gap}. The photon frequency is $\Omega=\omega_c$ (resonant) for (a) and (b) and $\Omega=0.65\,\omega_c$ (off-resonant) for (c) and (d). We use a dimensionless flux $\zeta=0.003$ and $z=0.025$. The bands of the static system (red dashed lines) are also shown for comparison.

All features described in the previous section using the Dirac approximation are observed here. In particular,  there are dynamical gaps  at around $\varepsilon=\pm \hbar\Omega/2$ with a magnitude in agreement with Eq.~\eqref{dynamical_gap}. 
In Fig.~\ref{zigzag_ribbon_Floquet}(c) the degeneracy between the two flat modes at $\varepsilon=0$ is removed, one remains at $\varepsilon=0$ while the other shifts downwards. We identify the latter with the $n=0$ bulk Landau level, which obeys the selection rules Eq.~\eqref{selection_rules} and whose shifting is in good agreement with these. The other state corresponds to the edge state solution Eq.~\eqref{usymm_sol} given in Appendix~\ref{landau_levels_in_graphene}. Being an edge state, it does not couple to the bulk states and hence it is pinned  at $\varepsilon=0$.  A similar analysis applies to Fig.~\ref{zigzag_ribbon_Floquet}(a) except that the bulk $n=0$ Landau level here is split instead of shifted due to the resonance condition.

 \begin{figure}[t]
\begin{center}
\includegraphics[width=0.95\columnwidth]{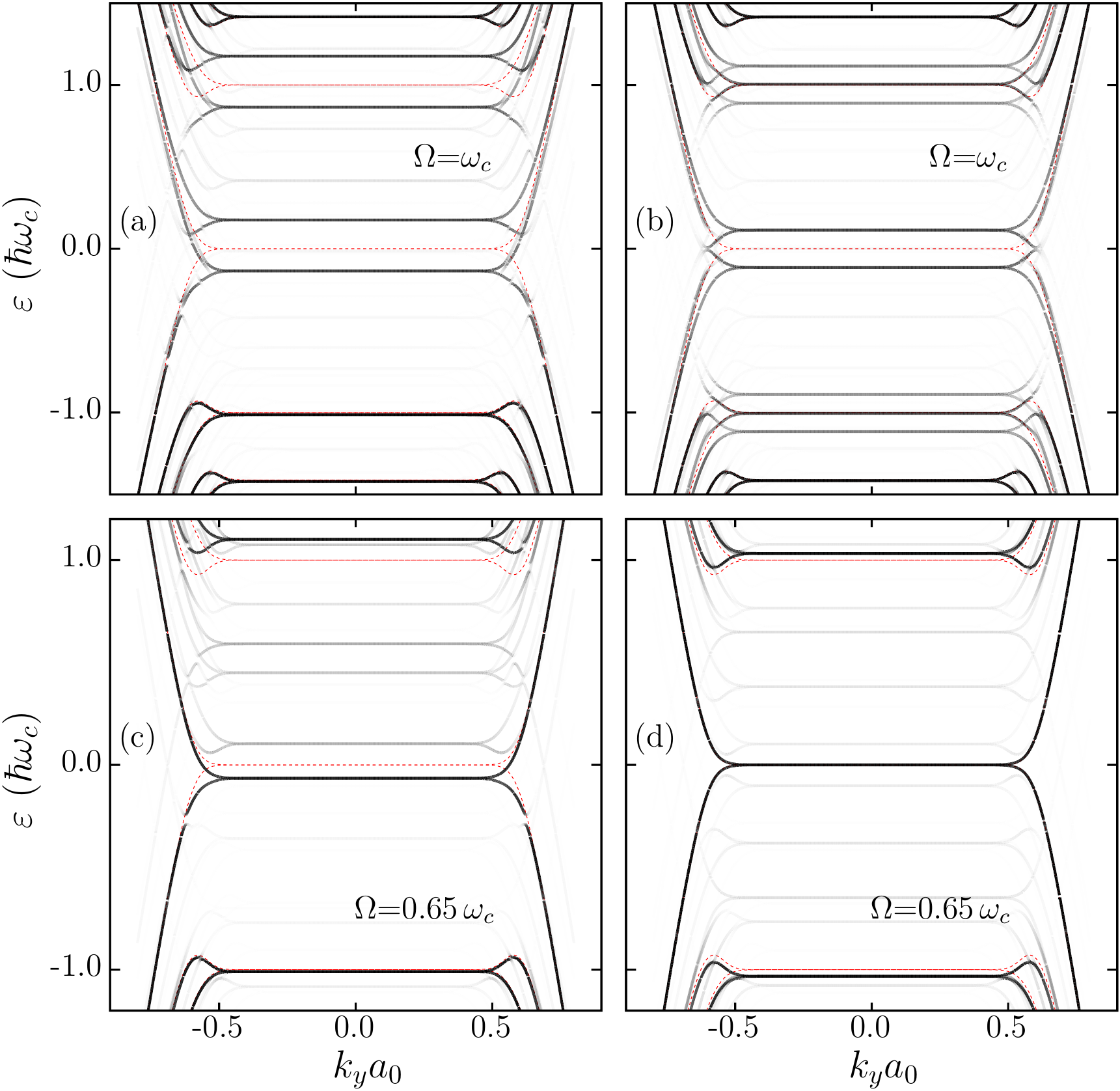}
\caption{(Color online) Same as previous figure but for an  armchair ribbon of width $W=120\sqrt{3}\,a_0$ (480 atoms in the unit cell). \label{armchair_ribbon_Floquet}}
\end{center}
\end{figure}
The linearly polarized case present similar features. However, in this case, the spectrum is electron-hole symmetric and hence the $n=0$ Landau level can only split.  For an off-resonant photon energy [Fig.~\ref{zigzag_ribbon_Floquet}(d)] the only effects of the laser field are the opening of the dynamical gaps at $\pm \hbar\Omega/2$ and the shifting of the first non-zero static Landau levels. In resonance, Fig.~\ref{zigzag_ribbon_Floquet}(b), there is a splitting in a neighborhood of $\varepsilon=0$. The states with $\varepsilon\neq 0$ are truly bulk states, coming from the mixing of the $n=0$ Landau level in the replica $m=0$, and the $n=\pm1$ Landau level from the replica $m=\mp1$. The state that remains at $\vre=0$ is the zigzag edge state mentioned above.

Figure \ref{armchair_ribbon_Floquet} shows the results for an armchair ribbon. The main feature predicted by the Dirac model is quite apparent: irrespective of the laser polarization, there is not a (first order) gap at $\pm\,\hbar\Omega/2$, the origin of such absence being the selection rule between  Landau edge states, Eq.~\eqref{armchair_selection_rule}.
Moreover, as it was mentioned in the preceding section,  the bands around $k_y=0$ (roughly $|k_ya_0|\leq 0.4$ in Fig.~\ref{armchair_ribbon_Floquet}), are basically bulk bands, and as such  their splitting follows the selection rules obtained in Sec.~\ref{bulk_selection_rules}.

\section{Two-terminal conductance\label{two_terminal_conductance}}
\begin{figure}[t]
\includegraphics[width=0.9\columnwidth]{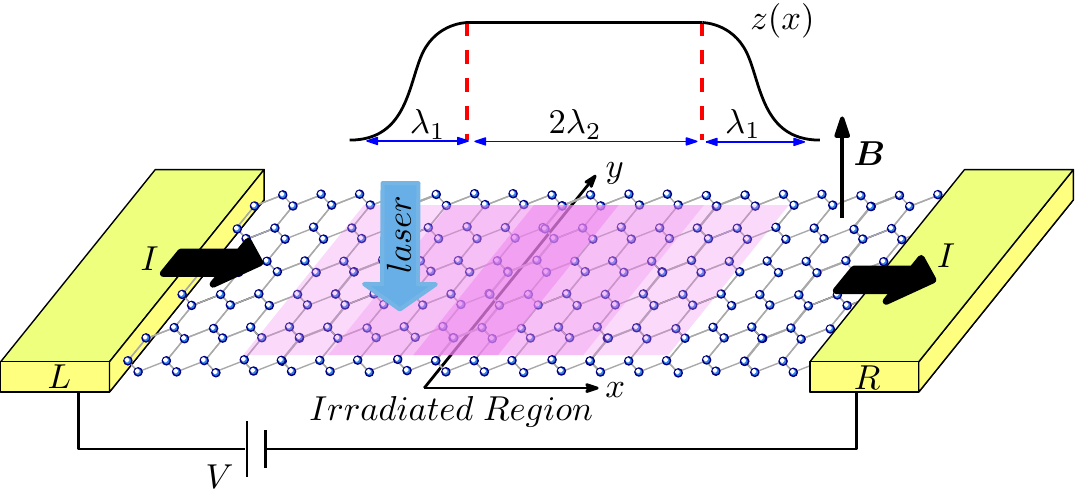}
\caption{(Color online) Setup for the calculation of the two-terminal conductance $G_{2T}$ using the Landauer-B\"uttiker approach for Floquet systems.
The laser field, defined by the local function $z(x)$, is applied along a central region and its  intensity  vanishes smoothly towards the leads\label{set-up}.}
\end{figure}

We now discuss the transport properties of  an illuminated ribbon in the QH regime in a two-terminal setup. For that, we take the magnetic field to be present throughout the entire sample (including the  semi-infinite leads), whereas the laser field is switched on smoothly (over a length scale $\lambda_1$),  kept constant for a distance $2\lambda_2$ and finally switched off, as schematically shown in Fig.~\ref{set-up}. This defines the scattering region. If we take the coordinate $x$ to be directed along the ribbon, then  the scattering region is defined by $|x|\leq \lambda_1+\lambda_2$, while the  local laser field intensity parameter $z(x)=ea_0A_0(x)/\hbar c$  is taken to be
\begin{equation}
z(x)=\left\{
\begin{array}{cl}
z\;, &|x|\leq \lambda_2 \\
\frac{z}{2}\left[1+\cos\left(\frac{\pi (|x|-\lambda_2)}{\lambda_1}\right)\right]\;, &\lambda_2\leq |x|\leq \lambda_1+\lambda_2\,.
\end{array}\right.
\end{equation} 
Here $z$ is the maximum value reached by the laser intensity. We use this symmetric profile for the laser field to preserve the left/right symmetry of the ribbon---the intensity is homogeneous along the transverse direction.

The current is computed within a scattering approach~\cite{moskalets_floquet_2002,camalet_current_2003}. In the non-interacting limit this is equivalent to the Keldysh formalism~\cite{arrachea_relation_2006,kohler_driven_2005}. This has been used for a variety of systems including laser illuminated graphene~\cite{foa_torres_multiterminal_2014,bajpai_robustness_2020}.
The time-average current, $\bar{I}=\frac{1}{T}\int_0^T dt\, I(t)$, is calculated according to
\begin{equation}\label{current1}
\bar{I}=\frac{2e}{h}\sum_n \int  \left[T_{RL}^{(n)}\,f_L(\varepsilon) - T_{LR}^{(n)}\,f_R(\varepsilon)\right]d\varepsilon\,,
\end{equation}
where $T_{RL}^{(n)}(\varepsilon)$ is the transmission probability for an electron with energy $\varepsilon$ from lead $L$ to lead $R$ emitting (absorbing) $n>0$ ($n<0$) photons and $f_{\alpha}(\varepsilon)$ is the Fermi function of the lead $\alpha$. Defining the quantities $T(\varepsilon)=\sum_n\,(T_{LR}^{(n)}(\varepsilon)+T_{RL}^{(n)}(\varepsilon))/2$ and $\delta T(\varepsilon)=\sum_n(T_{LR}^{(n)}(\varepsilon)-T_{RL}^{(n)}(\varepsilon))/2$, the average current $\bar{I}$ can be written as the sum of two terms
\begin{equation}\label{current_bias}
\bar{I}\!=\!\frac{2e}{h}\!\int\!  \left[ T(\varepsilon)(f_L(\varepsilon)\!-\!f_R(\varepsilon))\!-\!\delta T(\varepsilon)(f_L(\varepsilon)\!+\!f_R(\varepsilon))\right]\,d\varepsilon\,.
\end{equation}
At zero temperature, and up to first order in the bias difference $\delta V$, it reduces to
\begin{equation}\label{current_zero_T}
\bar{I}=\frac{2e^2}{h}T(\varepsilon_F)\,\delta V-\frac{4e}{h}\int_{-\infty}^{\varepsilon_F}  \delta T(\varepsilon)\,d\varepsilon\,.
\end{equation}
Here $\ef$ is the Fermi energy. The bias independent contribution in Eq.~\eqref{current_zero_T} is the so-called {\it pumped}  current. The inversion symmetry of our geometry guarantees that $\delta T(\varepsilon)=0$. We can then define the linear dc two-terminal conductance  $G_{2T}(\ef)=\bar{I}/\delta V=(2e^2/h)\, T(\ef)$ in terms of the transmittance at the Fermi energy. The latter is calculated using the Green function recursion technique within the Floquet formalism. 

\subsection{Zigzag ribbons}
We consider in this section a $300$ atoms wide zigzag ribbon ($W=224\,a_0$) and take $\lambda_1=2\lambda_2=800a_0\sqrt{3}$. This value is large enough as  to minimize the backscattering of electrons at the interface where the laser is on. 
The other parameters are $\zeta=0.003$ ($\hbar\omega_c\simeq0.46\,$eV and $\ell_B\simeq 13\,a_0$) and $z=0.025$. Unless otherwise mentioned, in all transport calculation we use only three Floquet replicas ($-1\leq m\leq1$),  with the aim of describing the most important features while reducing the computational cost. 
\begin{figure}[t]
\begin{center}
\includegraphics[width=0.95\columnwidth]{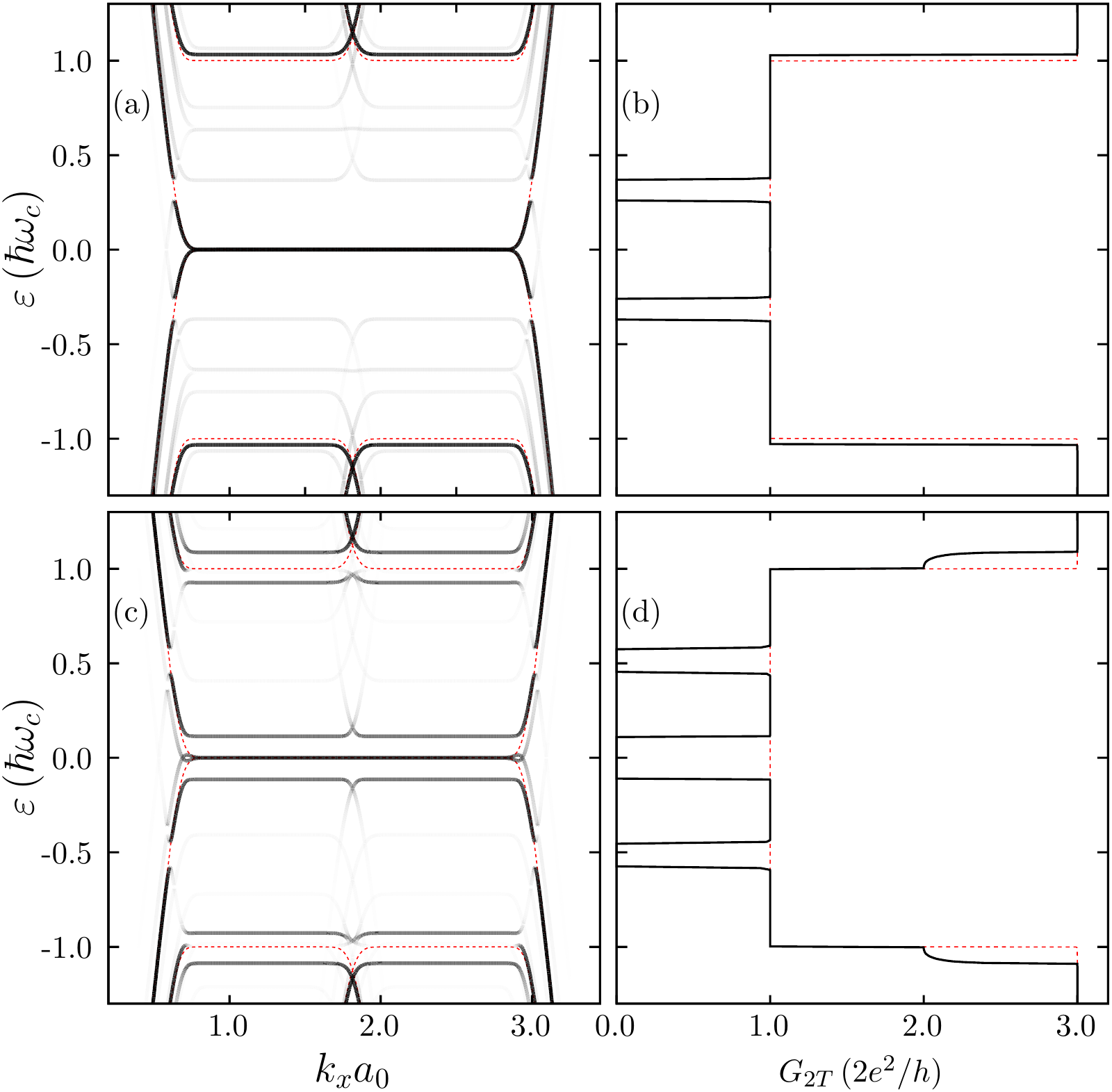}
\caption{(Color online) Landau-Floquet bands [(a) and (c)] and linear conductance $G_{2T}(\varepsilon)$ [(b) and (d)] of a zigzag ribbon of width $W{=}224\,a_0$ and $\zeta{=}0.003$. The laser is linearly polarized ($\alpha=\pi/2$, $\phi=0$) and we  have used $z{=}0.025$. Sub-plots (a) and (b) correspond to $\Omega{=}0.65\,\omega_c$, whereas in (c) and (d) we have $\Omega{=}\omega_c$ (resonance). The results of the static system are also included in red dashed lines for comparison.\label{2T_zigzag_linear}}
\end{center}
\end{figure}
First let us analyze the case of a linearly polarized laser ($\alpha=\pi/2$, $\varphi=0$). The results are shown in Fig.~\ref{2T_zigzag_linear} for the off-resonant (top panels) and resonant (bottom panels) situations. For each case, the conductance $G_{2T}$ [(b) and (d)] is shown by the side of the corresponding Landau-Floquet spectral density (projected  on the $m=0$ replica)  [(a) and (c)]. Here, electron-hole symmetry guarantees that  $G_{2T}(-\varepsilon)=G_{2T}(\varepsilon)$. 
 
For an off-resonant photon frequency [$\Omega=0.65\,\omega_c$,  Figs.~\ref{2T_zigzag_linear}(a) and~\ref{2T_zigzag_linear}(b)], two changes appear in the conductance (as compared with that of a non illuminated ribbon shown in red dashed lines): the two dynamical gaps centered at $\pm\,\hbar\Omega/2$, and a small energy shift in the $n=1$ (electron and hole) Landau levels where a transition from $G_{2T}=2e^2/h$ to $G_{2T}=6e^2/h$ takes place. 
In the quasienergy region corresponding to the dynamical gaps the conductance is almost completely suppressed in a very sharp way. This is due to the fact that transport is carried out entirely by edge states, which are completely reflected by the laser spot in that particular energy range owed to the appearance of a gap---a related effect was discussed in the case of a driven transition-metal dichalcogenide ribbon in Ref.~\cite{Huaman2019}.

When the photon energy is in resonance with the first non zero Landau level [$\Omega=\omega_c$, Figs.~\ref{2T_zigzag_linear}(c) and~\ref{2T_zigzag_linear}(d)], the conductance exhibits, in addition to the two dips at $\pm\,\hbar\Omega/2$, a strong suppression coming from the low energy gap created in the Floquet spectrum---it is interesting to note that the faint dispersive states near $\vre=0$ have a negligible contribution to the conductance.
Additionally, whereas in a non irradiated sample the conductance jumps from $2e^2/h$ to $6e^2/h$ when the Fermi energy crosses $\hbar\omega_c$ as the result of the change of the number of available edge states with a given chirality from one to three, here an intermediate quasi plateau at $G_{2T}\approx4e^2/h$ appears. 
As mentioned in the discussion of Fig.~\ref{zigzag_Floquet_spectrum}, this effect is related to the fact that there is a range of quasienergies were the number of effective edge modes is reduced by virtue of the upward energy shift of the Floquet edge mode of one of the valleys (the $K'$ valley in the case of Fig.~\ref{zigzag_Floquet_spectrum}). The origin of that shift is the level repulsion between the flat (dispersionless) edge state in the $m=1$ replica and the first dispersive ($\nu_2$) edge mode of the $m=0$ replica, which has always a higher quasienergy (this is not the case in the other valley). As such, this happens for opposite valleys in opposite sides of the sample. This explains why in Fig.~\ref{2T_zigzag_linear}(c) there are no edge modes for $\ef\sim \hbar \omega_c$ near $k_xa_0\sim1.8$ for the chosen ribbon's width. Due to the electron-hole symmetry, a similar argument holds for $\ef=-\hbar\omega_c$.
\begin{figure}[t]
\begin{center}
\includegraphics[width=0.95\columnwidth]{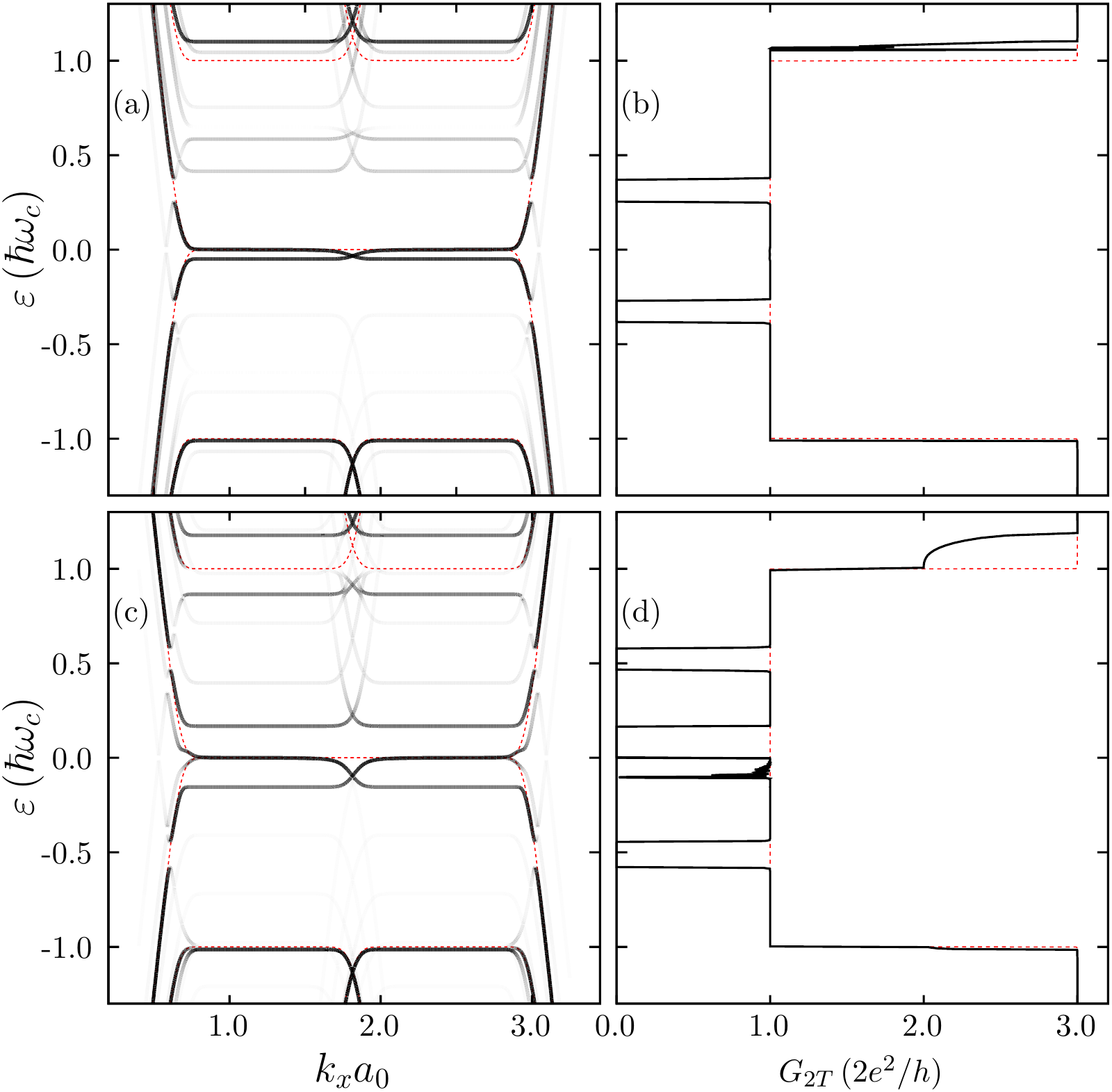}
\caption{(Color online) Same as Fig.~\ref{2T_zigzag_linear} but with a right-handed circularly polarized laser ($\varphi=2\alpha=\pi/2$).  The other parameters remain the same.
\label{2T_zigzag_circular}}
\end{center}
\end{figure}

The results for a circularly polarized laser ($\varphi=2\alpha=\pi/2$) are shown in Fig.~\ref{2T_zigzag_circular} for the same photon frequencies. All the prominent features described for the linear case are also observed here, but with the addition of several important new ones: 
\begin{enumerate}[label=(\roman*)]
\item In contrast to Figs.~\ref{2T_zigzag_linear}, here $G_{2T}(-\vre)\neq G_{2T}(\vre)$; 
\item When $\Omega=0.65\,\omega_c$ the conductance is, apart from the two square dips around $\pm\,\hbar\Omega/2$, quite similar to that of the static system. Notice that bending of the states near $\ef=0$ induced by the laser provides the channel that leads to $G_{2T}=2e^2/h$ near the Dirac point;
\item When $\Omega=\omega_c$, the emergence of the new dispersive edge modes near $\ef\sim0$ (on the negative side for our choice of polarization) is fully developed. These edge modes are the ones described in Fig.~\ref{zigzag_Floquet_spectrum} when discussing the properties of the edge modes of the $K'$ valley [and the same as those described above in (ii)]. They lead to a quantized conductance that partially fills the gap near zero energy---the narrow dip in the conductance that is observed in that energy region is due to a high-order anticrossing, which cannot be fully appreciated in the Landau-Floquet spectral density [Fig.~\ref{2T_zigzag_circular}(c)]. The quasi plateau in Fig.~\ref{2T_zigzag_circular}(d) slightly above $\ef=\hbar\omega_c$ has the same origin as those seen in Fig.~\ref{2T_zigzag_linear}(d), and can be explained in an analogous manner. The lack of a similar feature at $\ef=-\hbar\omega_c$ comes from the selection rules Eq.~\eqref{selection_rules}.
\end{enumerate}
\subsection{Armchair ribbons: the role of \textbf{\emph{adiabaticity}}}\label{armchair_ribbons_the_role_of_adiabaticity}
\begin{figure}[t]
\includegraphics[width=0.95\columnwidth]{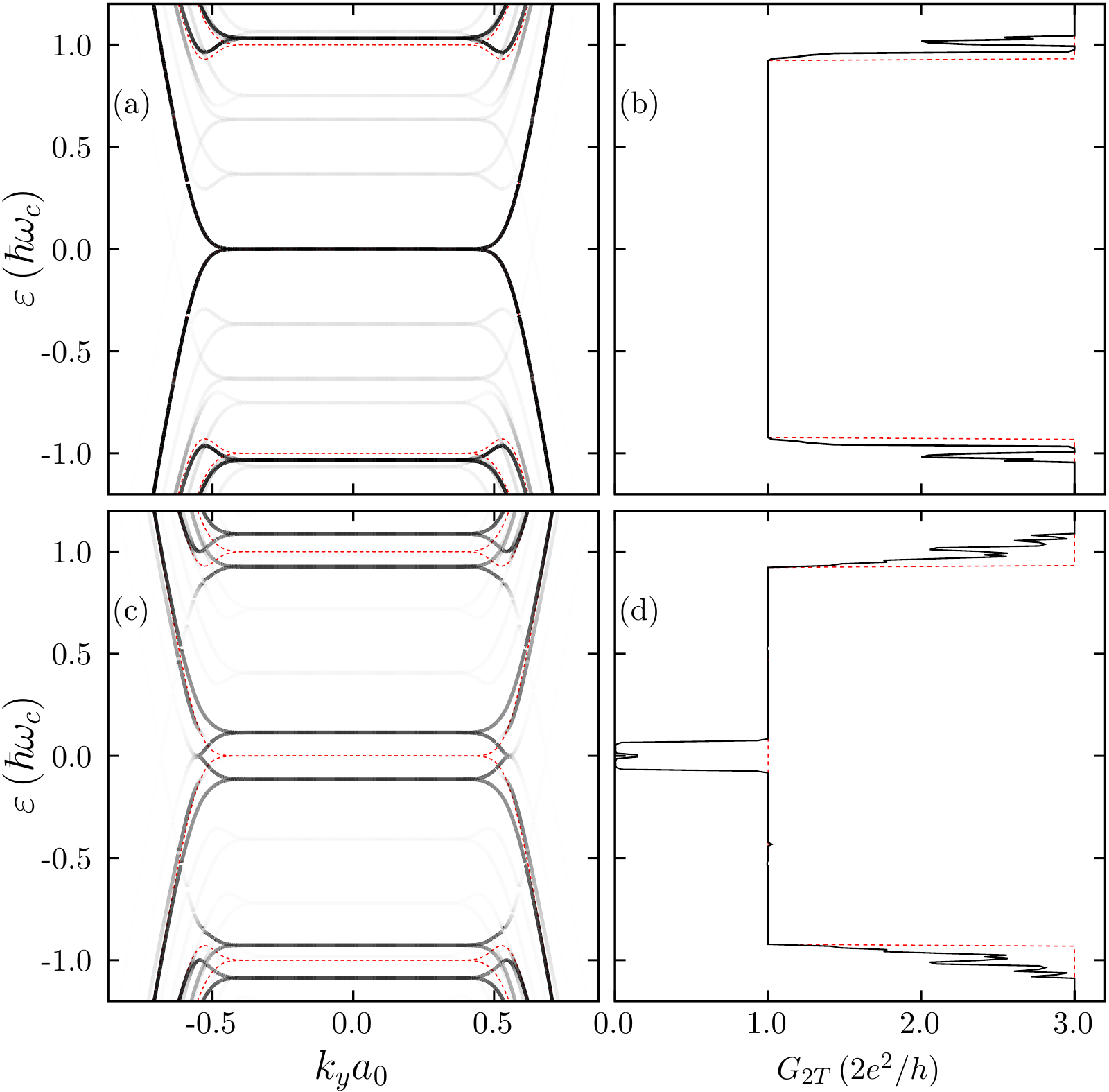}
\caption{(Color online) Landau-Floquet spectral density and two-terminal conductance for an armchair ribbon ($W=120\sqrt{3}\,a_0$)  under illumination with a linearly polarized ($\alpha=\phi=0$) laser. The parameters used are $\zeta=0.003$ and $z=0.025$. The photon frequency is $\Omega=0.65\,\omega_c$ [(a) and (b)] and $\Omega=\omega_c$ [(c) and (d)]. \label{armchair_linear_conductance}}
\end{figure}

We now consider an armchair ribbon of width $W=120\sqrt{3}\,a_0$ and keep the same parameters, $\zeta=0.003$ and $z=0.025$.  Figure  \ref{armchair_linear_conductance} shows the results for a linearly polarized laser ($\alpha=\varphi=0$, in-plane electric field normal to ribbon's edges). Let us point out first some general considerations. As always with this type of polarization, the Landau-Floquet spectrum presents electron-hole symmetry and thus  $G_{2T}(\ef)=G_{2T}(-\ef)$. Moreover, as the Landau-Floquet bands lack the dynamical gaps at $\pm\,\hbar\Omega/2$ (to first order in $z$), $G_{2T}$ does not show the typical strong suppression around these points---at most some minor very narrow features can be observed, corresponding to higher order photon processes. Additionally, for $\ef$ slightly below (above) $\hbar\omega_c$ ($-\hbar\omega_c$), where the static conductance changes  from $G_{2T}=2e^2/h$  to $G_{2T}=6e^2/h$, the conductance is rather oscillating, a behavior reminiscent of the $4e^2/h$ feature found in the zigzag case---note that, in the latter case, the features appear exactly at $\ef=\pm\,\hbar\omega_c$.
\begin{figure}[t]
\begin{center}
\includegraphics[width=0.95\columnwidth]{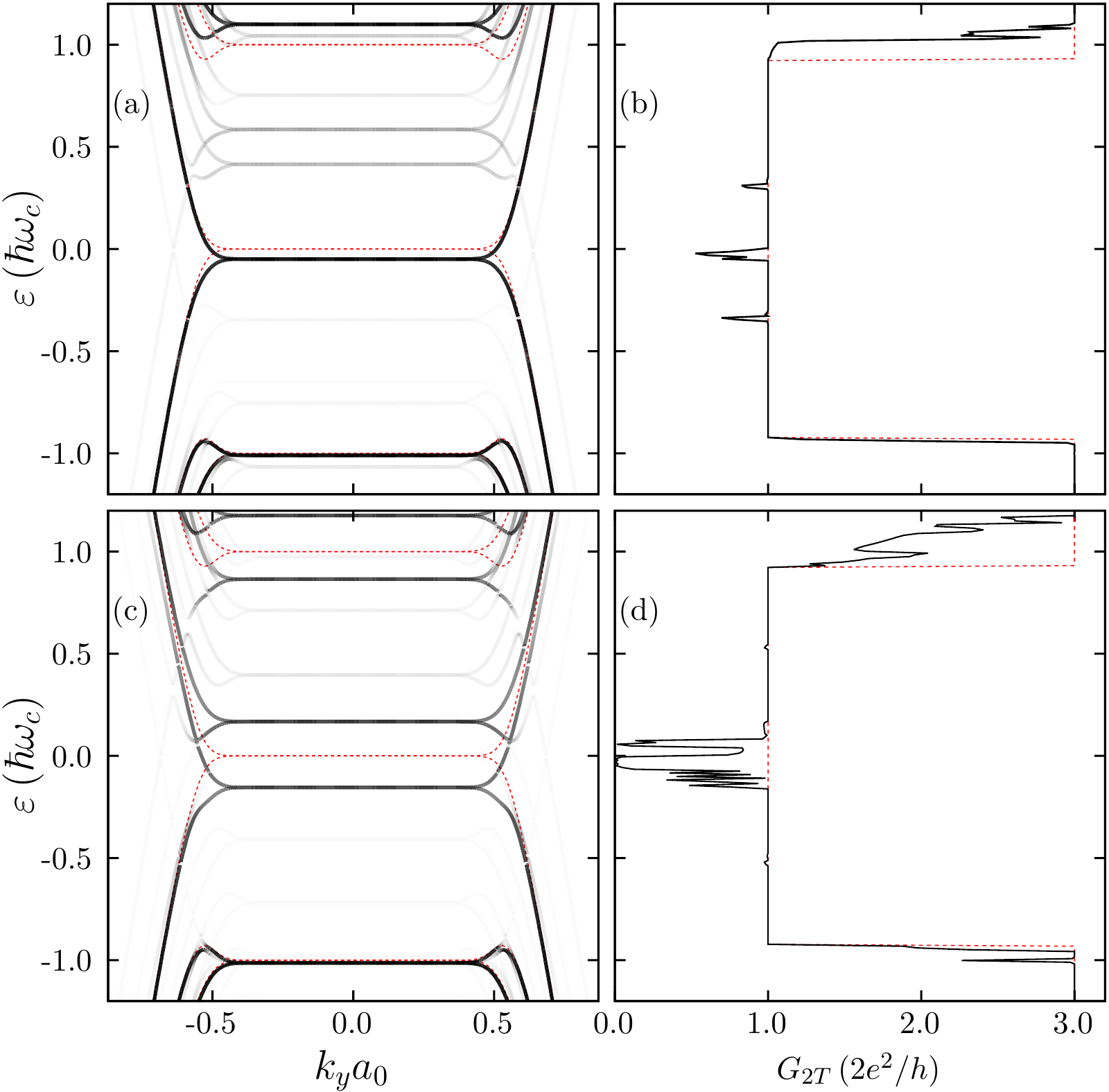}
\caption{(Color online) Same as Fig.~\ref{armchair_linear_conductance} but for a circularly polarized laser field  ($\varphi=2\alpha=\pi/2$). \label{2T_armchair_circular}}
\end{center}
\end{figure}

For a non-resonant photon frequency $\Omega=0.65\,\omega_c$, Figs.~\ref{armchair_linear_conductance}(a) and~(b), there are not special features around $\ef=0$. The linear conductance shows a flat profile as a function of $\ef$, nearly the same as the static system, in agreement with the small changes induced by the laser field on the spectral density. 

The situation changes when the laser photon frequency is in resonance, $\Omega=\omega_c$, Figs.~\ref{armchair_linear_conductance}(c) and~(d). On the one side, there is a sharp dip around $\ef=0$. Its limits are roughly defined by the small avoided crossings, coming from the coupling between the replicas $|\bm{\chi}_{\nu_1 k}^c,0\rangle$ and $|\bm{\chi}_{\nu_2 k}^v,1\rangle$ ($\ef>0$), and between $|\bm{\chi}_{\nu_1 k}^v,0\rangle$ and $|\bm{\chi}_{\nu_2 k}^c,-1\rangle$ ($\ef<0$)---selection rules Eq.~\eqref{armchair_selection_rule_general} state that the matrix element between these pairs of Floquet states is zero, so these gaps originate from higher order processes, which explains their smallness. 
This dip is not the product of an evanescent penetration inside the scattering region, since from Fig.~\ref{armchair_linear_conductance}(c) it is clear that there are conducting  states there. However, inside the region defined by the avoided crossings mentioned above, the Landau-Floquet states well inside the scattering region, and in a given edge of the ribbon, have the opposite sign of the velocity as compared with the incoming electrons. Therefore, the only way for these electrons to go through the central illuminated region is to move across the width of the ribbon until reach the opposite edge, where available states with a favorable velocity exist. If the laser's spatial profile (see Fig.~\ref{set-up}), is sufficiently smooth or {\it adiabatic}, as it is in our calculations, this motion of charge between edges is hindered by the presence of the small gaps introduced above. In this scenario, the fraction of incoming electrons reaching the opposite edge is negligible, and instead most of them simply backscatter into the Floquet channels $m=1$ or $m=-1$, depending on the character (conduction or valence) of the incident electrons. For electrons with energies just above the avoided crossing, the mismatch is still present, although in this case the incident electrons can reach the other edge and transmit into the other lead. This form of transmission is inherently inefficient, although it can be improved by further smoothing out the turning on-off of the laser (not shown). The renormalized (shifted) value of bulk part of $|\bm{\chi}_{\nu_2 k}^c,-1\rangle$ marks the onset of a constant  conductance $G_{2T}=2e^2/h$. Above this value there are two available Landau-Floquet channels, although with only one incident channel the transmittance reduces to $T(\ef)=1$. Due to electron-hole symmetry, this analysis can be extended to negative energies.
\begin{figure*}[t!]
\includegraphics[width=0.85\textwidth]{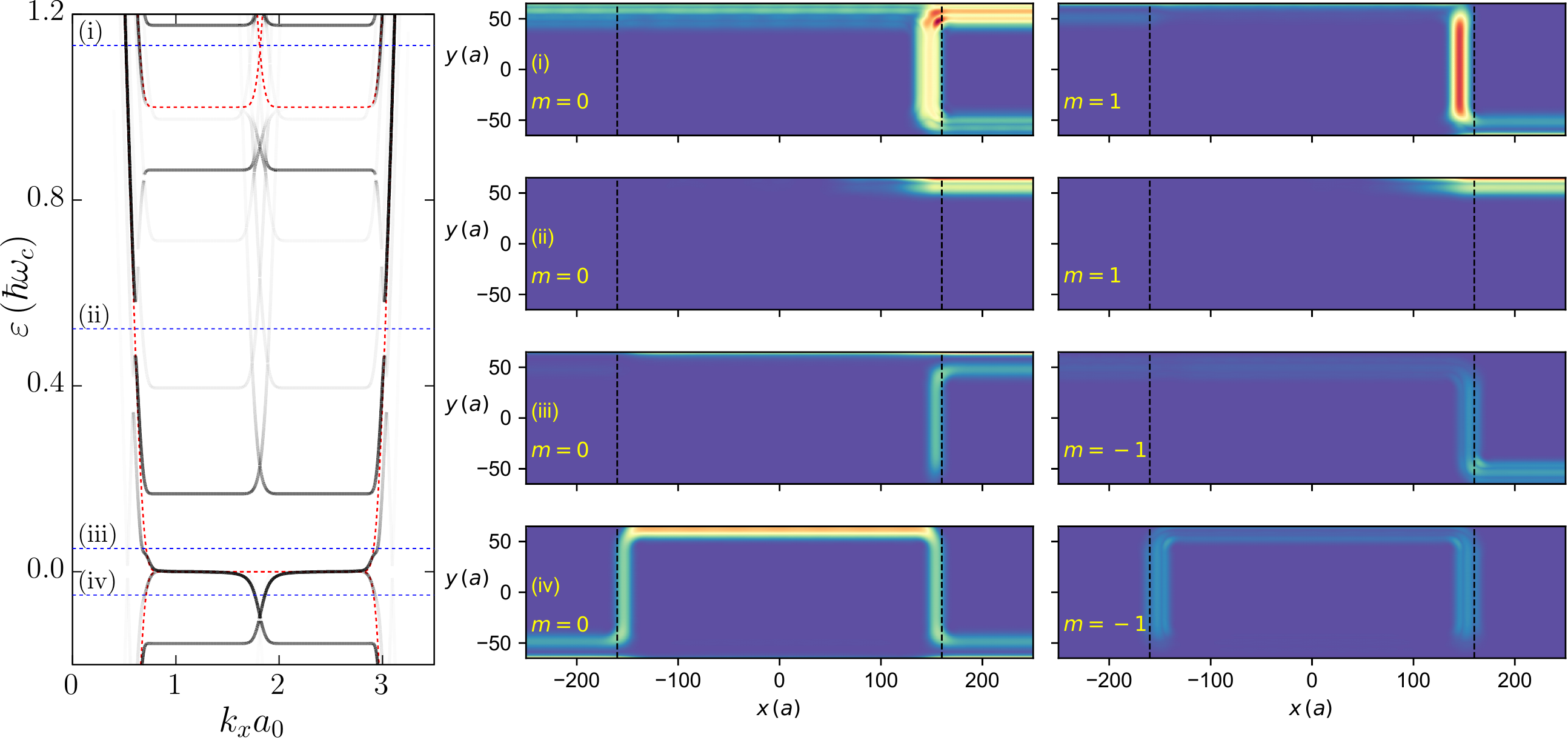}
\caption{(Color online) Scattering states as a function of the position in a zigzag ribbon for four selected values of the quasienergy. The width of the ribbon is $W=130\,a$. The parameters of the irradiated region (see Fig.~\ref{set-up}) are $\lambda_1=130\,a$ and $\lambda_2=30\,a$, where $a$ is the lattice parameter, $a=a_0\sqrt{3}$. The vertical dashed lines determine the central irradiated region. We use parameters $\zeta=0.003$, $z=0.025$, $\Omega=\omega_c$  and the polarization is circular and right-handed. Three Floquet replicas are used ($-1\leq m \leq 1$). We show the scattering states projected over the $m{=}0$ replica (central column), and those projections over a replica different from zero carrying most of the weight. In all cases the electrons come from the right lead with velocities to the left and in the replica $m{=}0$,  in the upper or lower edge depending on the quasienergy. 
\label{scatterig_wave_zigzag}}
\end{figure*}

On the other hand,  near $\ef=\hbar\omega_c$ (similar at $\ef=-\hbar\omega_c$), instead of the well defined change from $T(\ef)=1$ to $T(\ef)=3$ in the non driven system, there is a progressive increase from $1$ to $3$, something that resembles the zigzag case (where an intermediate step with $T(\ef)\simeq 2$ was found). 

The circularly polarized case (Fig.~\ref{2T_armchair_circular}, $\varphi=2\alpha=\pi/2$) presents similar features, with the expected lack of electron-hole symmetry. It is worth mentioning that the mismatch problem that leads to the suppression of the conductance for $\ef\simeq 0$ are already apparent in the non resonant case [Figs.~\ref{2T_armchair_circular}(a) and~(b)], as well as higher order narrow features. What is more, in resonance [Figs.~\ref{2T_armchair_circular}(c) and~(d), $\Omega=\omega_c$] the low energy conductance gap contains some fine structure inside it, which can be understood using the same arguments we introduced when dealing with the linearly polarized case. However, a better and more appealing way to analyse this is to  look at the scattering states in Floquet space, as we do in the next section.
\subsection{Scattering wave functions in Floquet space}
\begin{figure*}[t]
\includegraphics[width=0.85\textwidth]{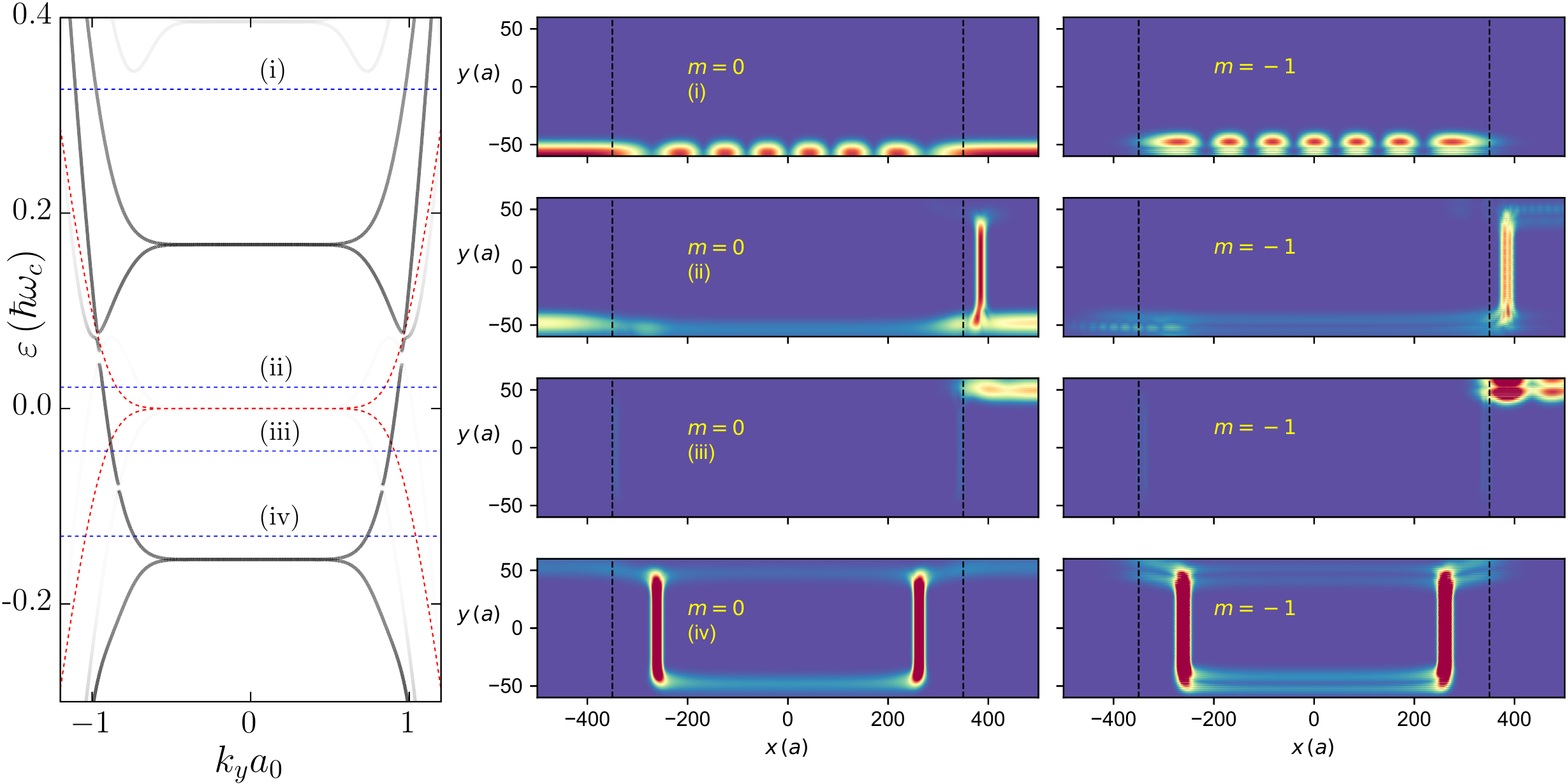}
\caption{(Color online) Same as the previous figure but for an armchair ribbon of width $W=120\,a$, $\lambda_1\approx 450\,a$ and $2\lambda_2\approx 200\,a$ ($a=\sqrt{3}a_0$). The parameters $\zeta$ and $z$, as well as the polarization of the laser, remain the same. The values of the quasienergies are: (i) $\vre=0.32\,\hbar\omega_c=0.15\,$eV, (ii) $\vre=0.022\,\hbar\omega_c=0.01\,$eV, (iii) $\vre=-0.043\,\hbar\omega_c=-0.02\,$eV and (iv) $\vre=-0.11\,\hbar\omega_c=-0.05\,$eV. For purposes of clarity, the saturation in the cases (iii) and (iv) has been enhanced.
\label{scatterig_wave_armchair}}
\end{figure*}

In the previous sections we described some of the features of the two-terminal dc conductance in terms of the appearance of gaps in the projected Landau-Floquet spectral density, the emergence of light induced edge states with different chirality, or the mismatch between the leads' and the system's wavefunction. All these effects can be made clearer by looking at the Floquet scattering states. Namely, we calculate the (squared) amplitude of the scattered wavefunction corresponding to an incident wave coming from the right lead, for instance, with a given energy in the $m=0$ Floquet replica.  The results are shown in Figs.~\ref{scatterig_wave_zigzag} and~\ref{scatterig_wave_armchair} for the zigzag and the armchair edge terminations, respectively. We have selected four particular values of the incident waves for illustration purposes.

In Fig.~\ref{scatterig_wave_zigzag} (zigzag ribbon, $W=130\,a$) the four selected scattered waves correspond to an incoming wave function with energy: (i)  $\vre/\hbar\omega_c=1.15$, in which case there are three propagating edge modes in the leads and hence the corresponding equilibrium conductance is $6e^2/h$. Of those edge modes, only two can propagate through the irradiated region, as it can be clearly observed in the projected Floquet spectral density shown in Fig.~\ref{scatterig_wave_zigzag}(i), while the other is fully reflected at the interface between the two regions (mostly in the $m=0$ replica on the other side of the sample). This leads to the quasi plateau of $4e^2/h$ shown in Fig.~\ref{2T_zigzag_circular}(d); (ii) $\vre=\hbar\Omega/2$, well inside the dynamical gap. Here the incident wave penetrates the scattering region as an evanescent wave before being reflected, on the same side of the sample, into the $m=1$ replica. Hence the conductance is strongly suppressed; (iii) $\vre/\hbar\omega_c=0.05$, inside the low energy Floquet gap but above the Dirac point. The incident wave is fully reflected on the $m=-1$ replica but on the other side of the sample; (iv) $\vre/\hbar\omega_c=-0.05$. Here the incoming wave, now on the bottom edge of the sample due to its valence band character,  is mostly fully transmitted  through the $m=0$ channel (with some participation of the $m=-1$ replica). However, quite remarkably, this requires the edge mode to \textit{switch} edges inside the irradiated region. That is, the Floquet edge mode inside the sample presents the opposite chirality as compared to the one it has on the (non irradiated) leads. As we show in the next section, this leads to a change in the sign of the Hall conductance.

Figure~\ref{scatterig_wave_armchair} shows the scattering states for an armchair ribbon of width $W=120\,a$. Here  we concentrate on a small quasienergy region, $\vre\in[-0.3\,\hbar\omega_c,0.4\,\hbar\omega_c]$, where  important departures from the static conductance appear [see Fig.~\ref{2T_armchair_circular}(c) and~(d)]. For all the quasienergies selected, the static system presents a conductance $G_{2T}$ equal to $2e^2/h$, with only one propagating channel coming from the right lead. In every case, we see that most of the incident flux is scattered through the channel $m=-1$, with a negligible component going into the channel $m=1$ (not shown here). This asymmetry in the roles of the Floquet replicas is a consequence of the time-reversal symmetry breaking imposed by the circularly polarized light and the selection rules Eq.~\eqref{selection_rules} and~\eqref{armchair_selection_rule_general}. Mainly due to the non monotonous bands of the second branch (see Appendix), this case shows some special peculiarities not present in zigzag ribbons.

As before, we consider four relevant cases with different energies: (i) $\vre/\hbar\omega_c=0.33$, there are two available channels inside the irradiated region, as the result of the superposition and coupling of two replicas $m=0$ and $m=-1$, for only one incident conduction channel. This leads to the appearance of oscillations in the probability density on each channel and to a displacement of the center of the orbits inside the illuminated region. The period of these oscillations is roughly equal to $2\pi/\delta k$, where $\delta k$ is the difference between the wave vectors $k_y$, corresponding to the incident energy $\vre$, of the two Floquet channels inside the sample.  The transmittance in this region is almost perfect so $G_{2T}\simeq 2e^2/h$; (ii) $\vre/\hbar\omega_c=0.022$, the sign of the velocity of the incoming channel matches that of the (only) Floquet states in the  irradiated region, and thus transport is possible, although imperfect due to the different spatial profile of both of them (incoming states are more centered towards the bulk while the Landau-Floquet states  are closer the edge). This results in a not very well developed peak of conductance [$T(\ef)\sim 0.9$], as it is shown in Fig.~\ref{2T_armchair_circular}(d).
In (iii) and (iv) the incoming channels move in the opposite direction to those available Landau-Floquet states inside the scattering region and on the same edge. Then, the only way for them to reach the other lead is to scatter into the other edge, where states with the same velocity are available for transport. Because of the adiabatic matching between the wavefunctions (as a consequence of the smooth turning on of the laser field), as it was already discussed in Sec.~\ref{armchair_ribbons_the_role_of_adiabaticity}, there is a certain energy threshold for this to happen set by the presence of a the small gap  between (iii) and (iv) as shown in the Floquet spectral density (see  Fig.~\ref{scatterig_wave_armchair}). Then we have the following: (iii) $\vre/\hbar\omega_c=-0.044$, here the incoming state cannot go through the avoided crossing and it is mostly reflected backward in the Floquet channel $m=-1$; (iv) $\vre/\hbar\omega_c=-0.11$, the the electrons arriving from the right lead can partially go through the avoided crossing in the Floquet spectrum, reaching the other edge of the ribbon.  Characteristic of this regimen is the noisy behavior of the conductance as a function of the Fermi level, leading to an incomplete transmission. 
Below $\vre=-0.07\,\hbar\omega_c$,  the velocities match and the transmission is perfect ($G_{2T}=2e^2/h$).
\begin{figure}[t]
\includegraphics[width=0.6\columnwidth]{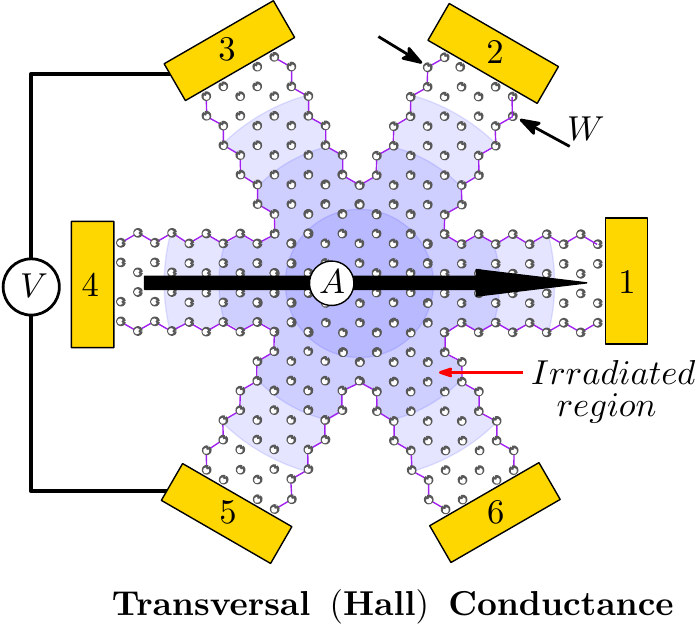}
\caption{(Color online) Zigzag six-terminal Hall bar to measure the Hall conductance  avoiding any pumped current. A similar setup can be established now with armchair leads.}\label{six_terminal_setup}
\end{figure}
\section{Hall conductance\label{six_terminal_conductance}}
In order to measure the Hall conductance one requires at least four terminals. However, the usual ($90^\circ$) four-terminal Hall bar configuration necessarily breaks the symmetry between leads, since if one pair of leads have zigzag edges, the other pair must have armchair edges. To avoid this, and to maintain the symmetry among all leads, we use the setup shown in Fig.~\ref{six_terminal_setup}. This six-terminal arrangement  has also the advantage that is does not generate any pumped current when the laser is turned on in a symmetrical  way, and hence it is simpler to obtain the conductance. Notwithstanding, the computational effort in calculating the scattering matrix \cite{Groth2014a} (and from this the Hall conductance) in this setup is more involved  when compared to the two terminal setup. 
\begin{figure*}[b!]
\includegraphics[width=1.\textwidth]{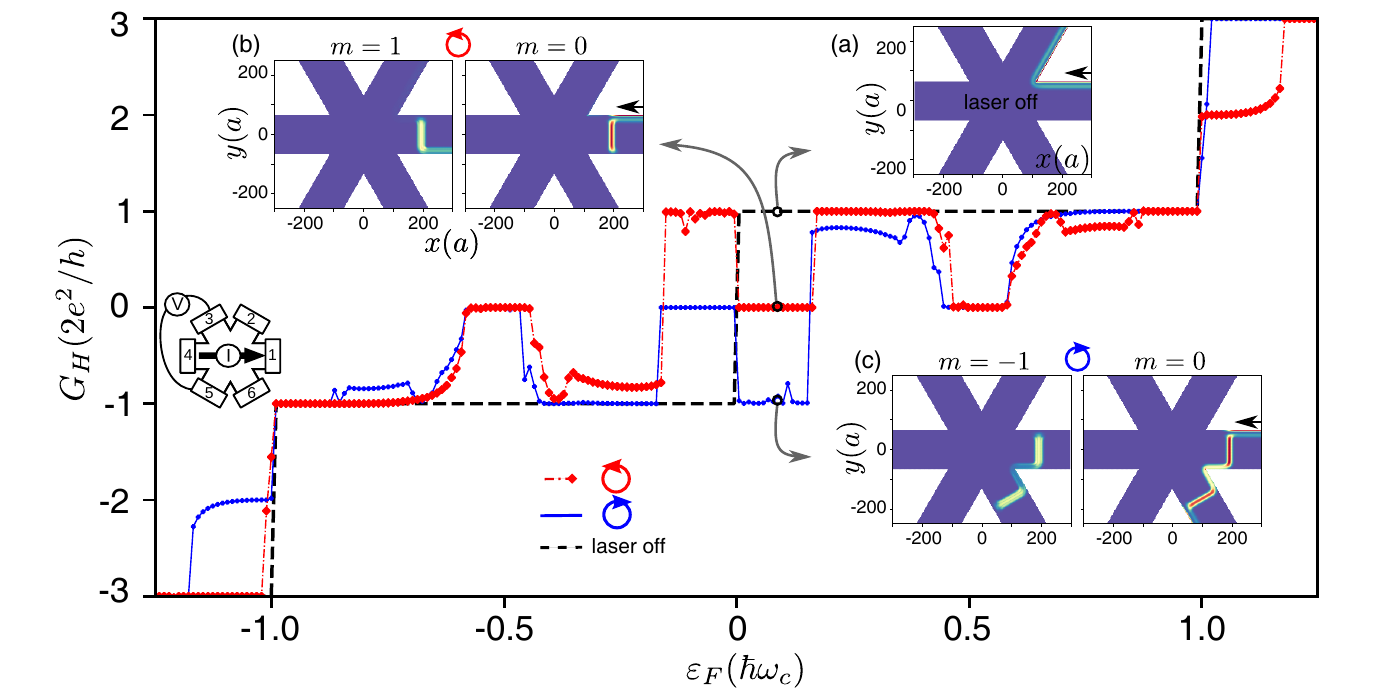}
\caption{(Color online) Hall conductance $G_H$ in a six-terminal setup (see inset) as a function of the Fermi level $\ef$. The width of the leads is $W=224\,a$ and the magnetic field corresponds to $\zeta=0.003$. The laser field is circularly polarized as indicated. The different insets shows the scattering states for the selected $\ef$ indicated in the main plot. In all cases the incoming state is in the channel $m=0$ (black arrow). Most of the scattering is into the $m=0$ channel and the $m=1$ ($m=-1$) for the ccw (cw) polarization. Both the switch-off of the conductance and the change of chirality are clearly seen for the ccw and cw polarization, respectively.\label{hall}}
\end{figure*}
Following Ref.~\cite{foa_torres_multiterminal_2014}, the time averaged current is now written as a generalization of Eq.~\eqref{current1}. If $\alpha$ and $\beta$ label  the terminals (leads),  the average current through the $\alpha$ lead is given by
\begin{equation}\label{landauer}
\bar{I}_\alpha=\frac{2e}{h}\sum_{\beta\neq\alpha}\sum_n\int \left[T^{(n)}_{\beta\alpha}(\varepsilon)f_{\alpha}(\varepsilon) - T^{(n)}_{\alpha\beta}(\varepsilon)f_{\beta}(\varepsilon)\right]d\varepsilon\,.
\end{equation}
The transmittances $T^{(n)}_{\alpha\beta}$ are the multi-terminal generalizations of those in Eq.~\eqref{current1}. Our six-terminal setup guarantees that $T^{(n)}_{\alpha\beta}$
is the same for any pair of adjacent terminals, ruling out the presence of any pumped current in the absence of a voltage bias.
In what follows, we assume that the chemical potential $\mu_\alpha$ at the lead $\alpha$ is not very different from  its equilibrium value $\ef$, that is,  $\mu_\alpha=\ef+\delta\mu_\alpha$ with $\delta\mu_\alpha$ small.
Taking the low temperature limit and expanding  Eq.~\eqref{landauer} up to linear terms in $\delta\mu_\alpha=-eV_\alpha$ we get to
\begin{equation}\label{landauer1}
\bar{I}_\alpha=\frac{2e^2}{h}\sum_{\beta\neq\alpha}\left[T_{\beta\alpha}(\ef)\,V_{\alpha} - T_{\alpha\beta}(\ef)\,V_{\beta}\right].
\end{equation}

Our goal is to study the Hall  conductance, as indicated in Fig.~\ref{six_terminal_setup}. To this we let a charge current flow from leads $4$ to $1$, $\bar I_4=-\bar I_1=I$, and impose $\bar I_\alpha=0$ on the remaining leads. From this we define the Hall conductance as $  G_H=I/(V_3-V_5)$. Furthermore, hereon we shall consider only the  circularly polarized resonant case, as it is the one with the most interesting features. 

Figure~\ref{hall} shows the Hall conductance for  a six-terminal system with  zigzag leads (width $W=130\,a$), as a function of the Fermi level $\ef$ for both chiralities of the circular polarization: $G_H^{\mathrm{ccw}}$ counterclockwise (ccw, \includegraphics[height=\fontcharht\font`\B]{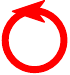}) and $G_H^{\mathrm{cw}}$ clockwise (cw, \includegraphics[height=\fontcharht\font`\B]{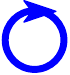})---the Hall conductance in the absence of a laser field is plotted in black dashed lines. The parameters are the same we have been using so far: $\zeta=0.003$ and $z=0.025$. Clearly, $G_H^{\mathrm{ccw}}(\ef)=-G_H^{\mathrm{cw}}(-\ef)$, as expected. Several features are apparent in the figure, which are directly related to those of the Floquet edge states discussed in the previous sections: (i) the suppression of the Hall signal around the dynamical gaps, $\ef\simeq\pm\,\hbar\Omega/2$, where the  scattering states only penetrate as evanescent waves inside the scattering region and hence the incoming states are almost fully backscattered; (ii) the presence of an intermediate plateau with $G_H\approx 4e^2/h$ near  $\ef=\hbar\omega_c$; (iii) an additional suppression for energies $\ef>0$ ($\ef<0$) near the Dirac point for the ccw (cw) case; and (iv) a switch of the sign the Hall conductance for energies right below (above) the gap mentioned in (c) for the ccw (cw) case. 

It is worth emphasizing that, quite remarkably, the switch of the Hall signal depends on the polarization of the laser field.  Furthermore, if we set $\ef$ to lay inside of the low energy gap [item (iii) above] for a given polarization, the Hall conductance can be turned on by simply changing the chirality of the polarization. This is depicted in the insets (b) and (c) of Fig.~\ref{hall} where we  show the scattering states for three selected energies. For the ccw polarization most of the incoming states  are backscattered into the same lead, which results in $G_H^{\mathrm{ccw}}\simeq0$. On the contrary, for the cw polarization, the incoming states cross over the width of the ribbon and reaches the other edge, from where it propagates to lead $6$, resulting in  a negative Hall signal. 

\begin{figure*}[t!]
\begin{center}
\includegraphics[width=1.0\textwidth]{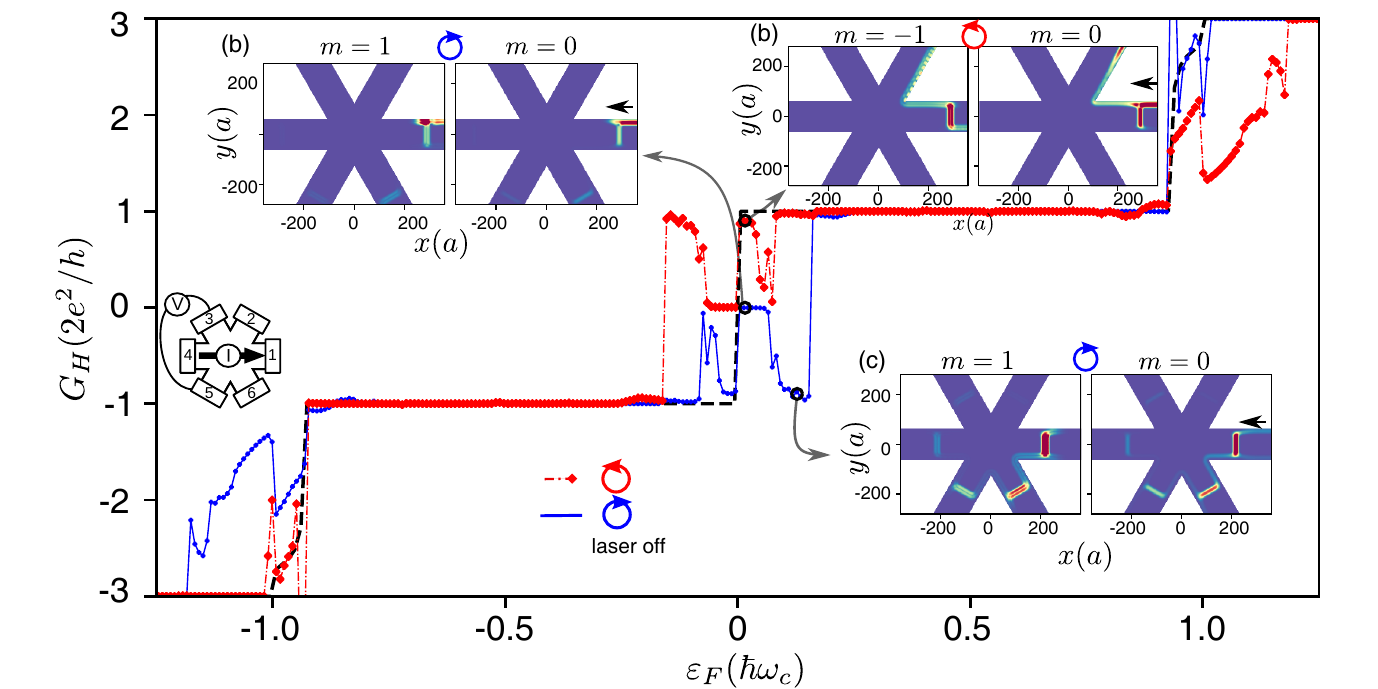}
\caption{(Color online) Same as in the previous figure but with armchair terminations (leads) of width $W=120\sqrt{3}\,a_0$. The insets correspond to quasienergies: (a) $\ef=0.01\,$eV, cw; (b) $\ef=0.01\,$eV, ccw;  (c) $\ef=0.06\,$eV, cw. The saturation has been enhanced for a better visualization of the incoming states, particularly in (c). \label{6T_armchair}}
\end{center}
\end{figure*}

When the leads have armchair edges, $G_H$ changes as shown in Fig.~\ref{6T_armchair}. Here, the parameters $\zeta$ and $z$ remain the same as in the zigzag case, while the width of the leads is now $W=120\,a$. 
First we notice that even in the absence of illumination (dashed black line), the armchair six-terminal geometry is accompanied by  a certain degree of back scattering. 
This effect takes place only in the $\ef$-range between the local minimum of the bands belonging to the second branch (see Appendix~\ref{landau_levels_in_graphene}), that in Fig.~\ref{6T_armchair}  are $|\bm{\chi}_{\nu_2k}^c\rangle$, and the corresponding bulk level, where counter propagating edge states do exist on the same side of the sample. This leads to the appearance of a shoulder-like structure just below $\ef=\hbar\omega_c$, that is  in the transition from $G_H=2e^2/h$ to $6e^2/h$. (The same is true near $\ef=-\hbar\omega_c$.)
This effect becomes less noticeable as the absolute value of the Landau bulk index $|n|$ increases.

When the laser is applied,  the departures from the static Hall conductance (black dashed line) take place in a small vicinity around the Dirac point, $\ef=0$, and around  $\pm\,\hbar\omega_c$, depending on the polarization (ccw or cw).   Let us describe what happens in a vicinity of $\ef=0$ ---as before, the scattering states for three selected energies, are shown as insets. Firstly, for positive values of $\ef$, $G_H^{\mathrm{ccw}}(\ef)$  exhibits a somewhat wide dip, indicating the presence of an avoided crossing in the spectrum. Apart from this, the Hall signal roughly follows its non driven value, as it can also be inferred from inset (b). On the other hand, $G_H^{\mathrm{ccw}}(\ef)$ vanishes just below $\ef=0$, and  changes sign for $\ef$ further below. The former behavior is analogous to that exhibit by the single ribbon, Sec.~\ref{armchair_ribbons_the_role_of_adiabaticity}: electrons hitting the illuminated region from lead $1$ (see inset on the far left), encounter counter propagating states there, but at the same time are unable to reach the opposite edge of the lead, where propagation would be possible, and simply backscatter. Inset (a) graphically describes the situation. In the situation where the Hall signal is reversed, the incident states reach the other side of the ribbon and spill into lead $6$, and the Hall signal passes from $-2e^2/h$ to $2e^2/h$. In the inset (c) the situation is shown for $G_H^{\mathrm{cw}}(\ef)$ and for a positive energy, although due to the symmetry $G_H^{\mathrm{ccw}}(\ef)=-G_H^{\mathrm{cw}}(-\ef)$ the result is entirely equivalent.

\section{Summary}\label{conclusions}
The quantum Hall effect is a paradigmatic topological phase, one of the most robust available phases, a fact that has proved useful for many applications. These developments came hand in hand with concrete measurements in different device setups. Here we present a study of the effect of strong laser illumination on graphene in the Quantum Hall regime and the changes in the Hall response in a multi-terminal setup that is accounted for by an atomistic description. Our results show that the quantum Hall plateaus can be disrupted or tailored by tuning the frequency, polarization and intensity of the laser field. This includes the switching on-off and a chirality inversion of the Hall signal, by tuning the handedness of the circular polarization.  

On one hand, the driving laser field induces the appearance of dynamical gaps in the Floquet spectrum that manifest, for instance, in the suppression of the Hall conductance near $\ef=\pm\,\hbar\Omega/2$---the fact that there is essentially a full suppression is related to the nature of the equilibrium transport in the QH regime which only occurs at the edges of the sample~\cite{Huaman2019}. This quench of the topologically protected QH transport arises from the resonant coupling between counter-propagating electron and hole edge states.
Quite interestingly, the $\pm\,\hbar\Omega/2$ gaps are absent in samples with  armchair edges and hence no suppression of the conductance is observed in such a case. This is a very particular aspect of the armchair termination, where the symmetry of the Hall edge modes leads to the cancellation of the dominant matrix elements of the time dependent perturbation. We have verified (but have not shown here) that the cove termination leads to similar results as the ones reported here for the zigzag case, and so we expect the suppression of the Hall signal at $\ef=\pm\,\hbar\Omega/2$  to represent the general situation. 
It is also worth mentioning that, under resonant conditions, and more clearly in the zigzag ribbon, the driving produces a novel $4e^2/h$ feature in the Hall conductance due to the reduction of the available edge channels near $\ef=\pm\,\hbar\omega_c$.

On the other hand, near the Dirac point, the driving becomes a non-equilibrium knob that allows to turn on and off the Hall signal or changing its sign. Quite remarkably, both effects depend on the polarization of the laser field.
The possibility to turn transport on and off relies on the appearance, under resonant conditions, of an effective transport gap right above or below the Dirac point, depending on the polarization's handedness. 
The change in sign of the Hall signal have a less anticipated and more striking origin. It results from a change of the chirality of the propagating modes inside the irradiated region that makes the propagating edge channel to cross from one side of the sample (ribbon or lead) to the other while inside the driving area.
In the zigzag case, for instance, the physical origin for this is the fact that the zero energy flat edge modes acquire a polarization-dependent dispersion that is always opposite to the one of the  regular QH edge states. 
We expect this to be a generic feature for all terminations where the $K$ and $K'$ valleys are separate and not fold onto each other. In this sense, the armchair case appears as the only exception.

We hope that the effects described in the present work will help in stimulating new experiments and theory on the interplay between chiral transport and periodic driving as this might open the door to new ways to control it. This includes systems in a QH phase as discussed here, but also topological insulators where spin-orbit coupling may add further intricacies~\cite{berdakin_spin-polarized_2020}.

\acknowledgements
We  acknowledge financial support from ANPCyT (grants PICTs 2016-0791 and 2018-01509), from CONICET (grant PIP 11220150100506), from SeCyT-UNCuyo (grant 2019 06/C603), FondeCyT (Chile) under grant number 1170917 and the EU Horizon 2020 research and innovation program under the Marie-Sklodowska-Curie Grant Agreement No. 873028 (HYDROTRONICS Project). GU and LEFFT acknowledge support from the ICTP associateship program and thank the Simons Foundation. LEFFT acknowledges discussions with Nikolai Kalugin, Paola Barbara and James McIver.

\appendix 
\section{Landau levels in graphene \label{landau_levels_in_graphene}}
For the sake of completeness we present here a brief  description of the Landau levels in graphene in the low energy approximation, which are obtained from Eq.~\eqref{Dirac_H} after the substitution $\bm{p}\rightarrow\bm{p}+\frac{e}{c}\bm{A}$.

Before starting, it is useful to point out the following symmetries of the hamiltonian $\Ha_0$,
\begin{eqnarray}
\nonumber
    \tau_x\otimes\sigma_y\,\Ha_0\,  \tau_x\otimes\sigma_y&=&\Ha_0\,,\\
        \tau_x\otimes\sigma_x\,\Ha_0\,  \tau_x\otimes\sigma_x&=&-\Ha_0\,,
         \label{symm}
\end{eqnarray}
that allow us to relate the eigenfunctions belonging to different valleys (before imposing the boundary condition) when the problem can be solved on each of them separately (see bellow).
\subsection{Bulk states}
In the Landau gauge, $\bm{A}(y)=-By\,\vs{x}$, the Hamiltonian Eq.~\eqref{Dirac_H} is invariant under translations in $x$ direction  so that the  eigenfunctions (for the $K$ valley)  can be written as $\Psi_n(x,y){=}L_x^{-\frac{1}{2}}\,e^{-\ci kx}\bm{\chi}_{nk}^K(y)$. Here $L_x$ is the samples's length along the $x$ direction and
\begin{equation}\label{bulk_function}
\bm{\chi}_{nk}^K(y)=\frac{1}{\sqrt{\ell_B\,(2-\delta_{n0})}}
\left(\begin{array}{c}
\phi_{|n|}(\tilde{y}) \\
\mbox{sgn}(n)\phi_{|n|-1}(\tilde{y})
\end{array}\right),
\end{equation}
where $\tilde{y}=y/\ell_B-k\ell_B$, $\ell_B=\sqrt{\hbar c/eB}$ is the {\it magnetic length}, and $\phi_n(x)$ is the normalized harmonic oscillator eigenfunction, $\phi_n(x)=(\sqrt{\pi}\,2^{n}n!)^{-1/2}\,e^{-x^2/2}H_n(x)$ with $H_n(x)$  the Hermite polynomial. The corresponding eigenvalues are $E_n{=}\mbox{sgn}(n)\hbar\omega_c\sqrt{|n|} $, where $n\in \mathbb{Z}$, sgn($n$) is the sign function  and $\omega_c=\sqrt{2}\vf/\ell_B$. Positive or negative values of $n$ correspond to electrons and holes respectively. Solutions for the $K'$ valley can be obtained by applying the operator $\sigma_y$ to Eq.~\eqref{bulk_function}, by virtue of Eqs.~\eqref{symm}.
\subsection{Zigzag edge}
The Landau levels corresponding to a semi-infinite plane with zigzag edges ending on a `B' site  can be obtained by requiring that the component `A' of the  wavefunction  be zero at the edge, which for the case depicted in Fig.~\ref{plane} corresponds to $y=0$---notice that (with the gauge $\bm{A}(y)=-By\,\vs{x}$) translational symmetry along the $x$ direction is preserved. Since this edge termination does not mix the two valleys $K$ and $K'$, the solutions can be found for each valley separately. These solutions still have the form $\Psi(x,y)=L_x^{-\frac{1}{2}}\mathrm{e}^{-\ci k x}\bm{\chi}_k(y)$ but the components of $\bm{\chi}_k(y)$ are not the harmonic oscillator eigenfunctions but the general solutions of the corresponding differential equations, and  that are well-behaved in the limit $y\rightarrow \infty$. These are the 
Parabolic Cylinder functions $D_\nu(x)$ with $\nu\in\mathbb{R}$ and $\nu\ge0$. Hence we have \cite{Gusynin2008}
 \begin{eqnarray}
 \bm{\chi}^K_{\nu k}(y)&=&\frac{1}{\sqrt{C_{\nu k}}} \left(
 \begin{array}{c}
 D_{\nu}(\xi) \\ \frac{\varepsilon}{\varepsilon_0} D_{\nu-1}(\xi)
 \end{array}
 \right),\label{basisK}\\
 \nonumber
 &&\\
 \bm{\chi}^{K'}_{\nu k}(y)&=&\frac{1}{\sqrt{C_{\nu k}}} \left(
 \begin{array}{c}
 \frac{\varepsilon}{\varepsilon_0}D_{\nu-1}(\xi) \\ -D_{\nu}(\xi)
 \end{array}
 \right),\label{basisKp}
\end{eqnarray}
where $\xi{=}\sqrt{2}(y/\ell_B-k\ell_B)$ and $C_{\nu k}$ is a normalization constant. The quantity $kl_B$ can be considered the {\it center} of the cyclotron orbit of the electrons, and thus the larger $kl_B$, the deeper into the bulk the electrons reside. The quantity  $\varepsilon=s\, \hbar\omega_c \sqrt{\nu}$ is the energy of the state ($s=\pm1$ refers to the electron and hole bands respectively),  with  $\omega_c{=}\sqrt{2}\vf/\ell_B$. 

\begin{figure}[t]
   \includegraphics[width=0.9\columnwidth]{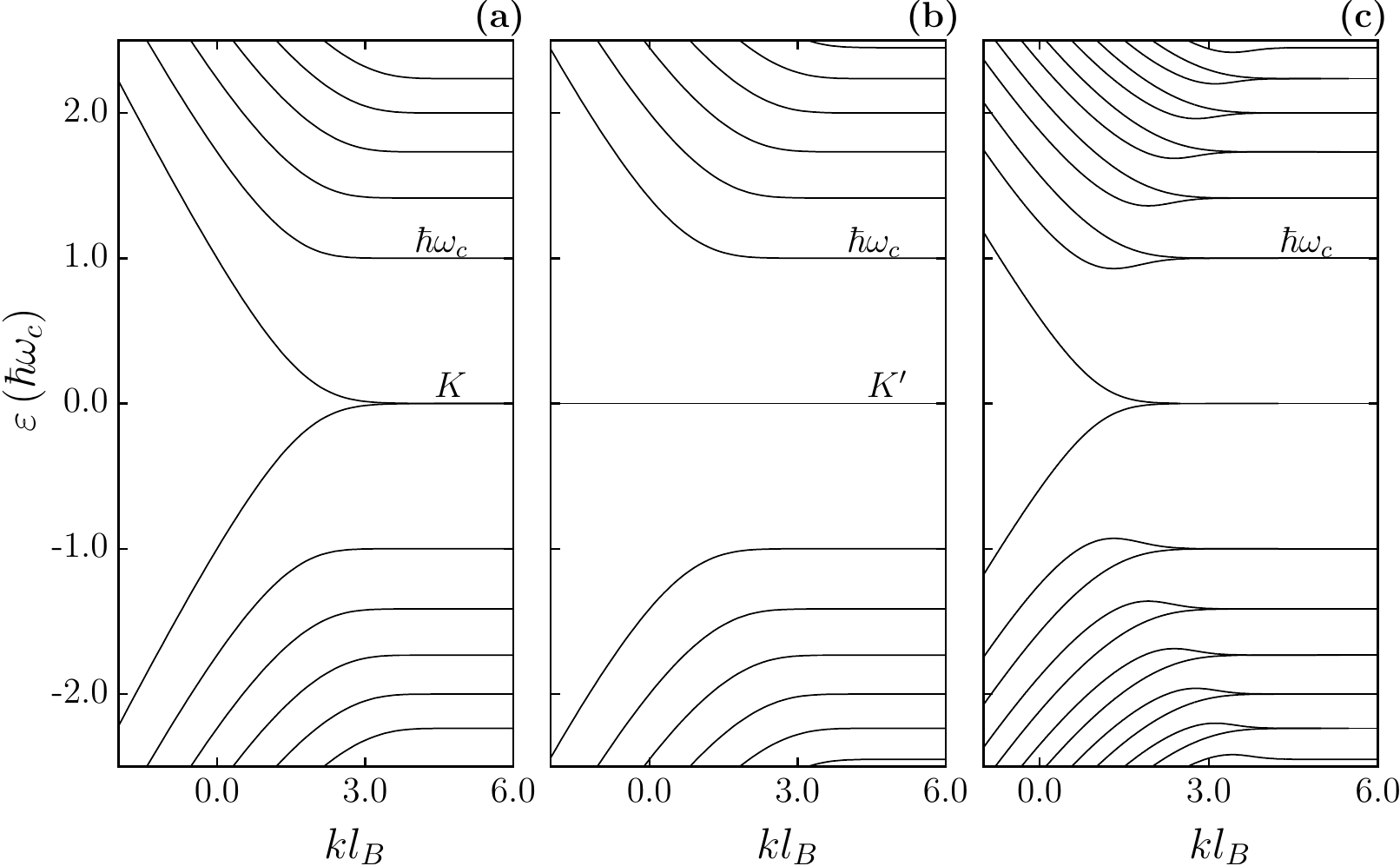}
   \caption{Energy dispersion (in units of $\hbar\omega_c$)  for the $K$ [(a)] and $K'$ [(b)] Dirac valleys in a zigzag edge. Note that the $K'$-valley have a dispersionless mode with $\varepsilon(k)=0$ while the $K$-valley possesses two states that converge to the $n=0$ bulk Landau level.  In [(c)] we show the energy dispersion for an armchair ribbon. No distinction between valleys has place here. In every case $\zeta=0.003$. The first non zero bulk Landau level $E_1=\hbar\omega_c$ is indicated.}
  \label{zigzag_dispersion}
\end{figure}

The index $\nu(k)$, and thus the energy dispersion, is determined by the boundary condition that the upper component (sublattice $A$) of the spinor wavefunction vanish at position $y=0$. That is
\begin{equation}
D_\nu(-\sqrt{2}k\ell_B)=0\,,\qquad \sqrt{\nu}\, D_{\nu-1}(-\sqrt{2}k\ell_B)=0,
\label{boundary_cond}
\end{equation} 
for the $K$ and $K'$ valleys, respectively. This gives a discrete set of eigenenergies $\varepsilon_l^K(k)$ and $\varepsilon_l^{K'}(k)$ for each value of $k$, where the integer $l$ labels the solutions of Eqs. \eqref{boundary_cond}.
It is straightforward to verify that because for the same value of $\nu_l$ the two spinors in the conduction and valence bands (opposite $s$)  must be orthonormal, one has that 
 \begin{equation}\label{bedingung}
     \int_{-\sqrt{2}k\ell_B}^{+\infty}d\xi\,D_{\nu_l}^2(\xi)=\int_{-\sqrt{2}k\ell_B}^{+\infty}d\xi\,\nu_l D_{\nu_l-1}^2(\xi)\,,
 \end{equation}
and hence the normalization constant becomes
\begin{equation}
C_{\nu_l k}=\sqrt{2}\ell_B\int_{-\sqrt{2}k\ell_B}^\infty d\xi\, D_{\nu_l}^2(\xi)\,.
\end{equation}
Figures~\ref{zigzag_dispersion}(a) and (b) show the energy dispersion for both valleys where dispersive edge states are apparent. Far from the boundary ($k\ell_B\gg1$) one has that $\nu_l\rightarrow n\in \mathbb{N}
^0$, and so  the bulk Landau levels $\pm\,\hbar\omega_c \sqrt{n}$ are recovered. Note that the $K'$ valley possesses dispersionless states pinned at the Dirac point, the corresponding spinor being the same as for the zero energy bulk Landau level.

The edge states in the $K$ valley close to zero energy deserve a separate and careful treatment. From Fig.~\ref{zigzag_dispersion}(a), it is clear that the lowest conduction and the highest valence bands converge into each other and into $\varepsilon=0$ as $k\ell_B$ becomes large. At first sight, it might appear, since in this case $\nu\rightarrow 0$, that the two spinor states themselves go into the same spinor proportional to  $[D_0(\xi),0]
^T$, with a vanishing second ($B$) component. This state clearly could be identified with the bulk $n=0$ Landau level [cf. Eq.~\eqref{bulk_function} with $n=0$ and $\phi_0(z)=D_0(z)$]. However, Eq.~\eqref{bedingung} shows that even in this limit ($\nu\approx 0$), the $B$-component does not vanish as the left side of the equation converges to $\sqrt{2\pi}$ when $k\ell_B\rightarrow\infty$.
To properly account for this one needs to take into account the precise way in which $\nu\rightarrow0$ as $k\ell_B\rightarrow\infty$. After expanding the equation $D_\nu(-\sqrt{2}k\ell_B)=0$ for $k\ell_B\gg1$, the solution for  vanishing $\nu$ is given by $\nu\simeq \pi^{-\frac{1}{2}}k\ell_B \exp{[-(k\ell_B)^2]}$ and therefore the $B$-component of the spinor wavefunction at the edge goes as $\sqrt{\nu}D_{\nu-1}(\xi)\simeq \pi^{\frac{1}{4}}(2k\ell_B)^{\frac{1}{2}}e^{-ky}            $. That is, the $B$-component gets sharper as $k\ell_B$ grows and so it is more localized at the edge (recall that the condition Eq.~\eqref{bedingung} must hold).
In summary, for  $k\ell_B\gg1$ the two orthonormal eigenstates have the $A$-component located mostly in bulk, and the $B$-component at the edge. The fact that $\nu$ is almost zero allow us to treat them as nearly degenerate, and thus change the basis to a symmetrical $\bm{\chi}_{\mathrm{bulk}}$ and antisymmetrical $\bm{\chi}_{\mathrm{edge}}$ solutions,
 \begin{eqnarray}
 \bm{\chi}_{\mathrm{bulk}}(y)&=&\sqrt{\frac{2}{C_{\nu k}}} \left(
 \begin{array}{c}
 D_{\nu}(\xi) \\ 0
 \end{array}
 \right),\label{symm_sol}\\ 
 \bm{\chi}_{\mathrm{edge}}(y)&=&\sqrt{\frac{2}{C_{\nu k}}} \left(
 \begin{array}{c}
 0\\ \sqrt{\nu} D_{\nu-1}(\xi)
 \end{array}
 \right)\,.\label{usymm_sol}
\end{eqnarray}
In this representation the spatial profile of each spinor is evident. As it was already pointed out, $\bm{\chi}_{\mathrm{bulk}}$ is to be identified with the $n=0$ bulk Landau level, whereas $\bm{\chi}_{\mathrm{edge}}$ is the well known non-dispersive edge state in a zigzag graphene ribbon~\cite{Nakada1996,Gusynin2008}.

\subsection{Armchair edge}
To be consistent with the geometry shown in Figs.~\ref{plane} and~\ref{lattice2}, we take the armchair edge to be along the $x=0$ line and therefore, to make use of the  translation symmetry along that direction, we use a different but completely physically equivalent gauge, namely  $\bm{A}(x){=}Bx\,\vs{y}$. The wavefunction on each valley can then be written as $L_y^{-1/2}\,e^{\ci ky}\bm{\chi}^{K(K')}_{\nu k}(x)$, where the spinor components are given by 
\begin{eqnarray}\label{basisa}
 \bm{\chi}^K_{\nu k}(x)&=&\frac{1}{\sqrt{\tilde{C}_{\nu k}}} \left(
 \begin{array}{c}
 \,D_{\nu}(\tilde{\xi}) \\ -\ci\frac{\varepsilon}{\varepsilon_0} D_{\nu-1}(\tilde{\xi})
 \end{array}
 \right),\\
 \bm{\chi}^{K'}_{\nu k}(x)&=&\frac{1}{\sqrt{\tilde{C}_{\nu k}}} \left(
 \begin{array}{c}
 \ci\frac{\varepsilon}{\varepsilon_0}D_{\nu-1}(\tilde{\xi}) \\  D_{\nu}(\tilde{\xi})
 \end{array}
 \right).
\end{eqnarray}
Here $\tilde{\xi}=\sqrt{2}(x/\ell_B-k\ell_B)$ and $\varepsilon=s \hbar\omega_c\sqrt{\nu}$. 
If we denote the complete wavefunction of the armchair edge as $L_y^{-1/2}\,e^{\ci ky}\bm{\chi}_{\nu k}(x)$, then the boundary condition in the armchair edge corresponds to make $\bm{\chi}_{\nu k}(0)=0$. This cannot be made separately in each valley (the boundary mixes the valley index) and so the solution needs to be constructed using a  linear combinations of both valleys~\cite{Brey2006a,Gusynin2008}
\begin{equation}\label{whole_eq}
\bm{\chi}_{\nu k}(x)=
\alpha\, e^{\ci Kx}\,\bm{\chi}^{K}_{\nu k}(x) + 
 \beta\, e^{-\ci Kx}\,\bm{\chi}^{K'}_{\nu k}(x)\,,
\end{equation}
where we have added a phase factor $e^{i\bm{K}\cdot\bm{r}}$ ($e^{i\bm{K}'\cdot\bm{r}}$)  to the $K$ ($K'$) component, noticing that in our geometry (see Fig.~\ref{lattice2}) we have $\bm{K}'=-\bm{K}=-K\vs{x}$, with $K=4\pi/(3\sqrt{3}a)$. The condition $\bm{\chi}_{\nu k}(0)=0$ results in the following equation for $\nu(k)$
\begin{equation}\label{armchair_cond}
D_\nu^2(-\sqrt{2}k\ell_B)-\nu\,D_{\nu-1}^2(-\sqrt{2}k\ell_B)=0\,.
\end{equation}
This equation yields a discrete infinite set of solutions for each $k$, that in ascending value  we denote as $\nu_n(k)$, with $n=1,2,\cdots$. It is interesting to note that these solutions can be arranged in a way that simplifies the construction of their corresponding eigenstates. To this, we first note that Eq.~\eqref{armchair_cond} can be written as follows (all $D_\nu$ functions are evaluated at $-\sqrt{2}k\ell_B$ )
\begin{equation}\label{armchair_cond_factorized}
(D_\nu-\sqrt{\nu}\,D_{\nu-1})(D_\nu+\sqrt{\nu}\,D_{\nu-1})=0\,.
\end{equation}
It can be shown that solutions $\nu_n$ with $n$ odd satisfy $D_\nu{-}\sqrt{\nu}\,D_{\nu-1}{=}0$, while those with $n$ even are solutions of $D_\nu{+}\sqrt{\nu}\,D_{\nu-1}{=}0$. To make things more compact, we introduce an index $\tau_n=(-1)^{n+1}$.
We then say that the $\nu_n$ solutions exist in two {\it branches} defined by the equation $D_{\nu_n}(-\sqrt{2}k\ell_B)=\tau_n\sqrt{\nu_n}\,D_{\nu_n-1}(-\sqrt{2}k\ell_B)$. 

The energy spectrum $\varepsilon_n^s(k)=s\,\hbar\omega_c\sqrt{\nu_n(k)}$ ($s=\pm1$) is shown in Fig.~\ref{zigzag_dispersion}(c). The bands with $n$ odd (the {\it first branch }), are monotonous decreasing functions of $k\ell_B$, very similar to those  of a zigzag edge at the $K$ point. For $n$ even (the {\it second branch}), the bands are decreasing functions for very negative values of $k\ell_B$, until they reach a local minimum and  start to grow as $k\ell_B$ increases. For $k\ell_B$ large enough ($k\ell_B>3$ if we consider the first six bands), we see that adjacent pairs of bands converge to each other and into the bulk states: $\{\nu_n,\nu_{n+1}\}\rightarrow\sqrt{n/2}$, with $n=2,4,6,\cdots$.

We now examine the specific structure of the corresponding  eigenfunctions as it plays a crucial role in determining the value of the gaps in the Floquet spectrum. To this, we first notice that the coefficients $\alpha$ and $\beta$ in Eq.~\eqref{whole_eq} satisfy $\alpha_n=-\ci \beta_n s\tau_n$.
With this result it is useful to write down these eigenfunctions in the four-component spinor notation,
\begin{equation}\label{gen_sol}
\bm{\chi}_{\nu_n k}^s(x)=\frac{1}{\sqrt{C_{\nu_n k}}}
\left[\begin{array}{c}
-\ci s \tau_n D_{\nu_n}(\tilde{\xi})e^{\ci Kx} \\
-\tau_n\sqrt{\nu_n} D_{\nu_n-1}(\tilde{\xi})e^{\ci Kx} \\
\ci s\sqrt{\nu_n} D_{\nu_n-1}(\tilde{\xi})e^{-\ci Kx} \\
 D_{\nu_n}(\tilde{\xi})e^{-\ci Kx}
\end{array}\right]\,,
\end{equation}
where the $s$ parameter indicates electron/conduction ($s{=}1$) or hole/valence ($s{=}-1$) character, which in the main text will be often denoted by superscripts $c$ and $v$, respectively. Since the probability density is obtained by summing up sub-lattice ($A$ and $B$) components, it can be shown that the spatial profile of $\bm{\chi}_{\nu_n k}^s(x)$ across the ribbon shows oscillations of the form $\cos2Kx$, with a period equal to $1/2K$ (the inverse of the distance between the two non equivalent valleys~\citep{Brey2006}). The normalization constant $C_{\nu k}$ is given by
\begin{widetext}
\begin{equation}
C_{\nu_n k}=\sqrt{2}\,\ell_B\int_{-\sqrt{2}kl_B}^{+\infty}\mbox{d}\xi\,[D_{\nu_n}^2(\xi)+\nu_n D_{\nu_n-1}^2(\xi)-\tau_n\sqrt{\nu_n}D_{\nu_n}(\xi)D_{\nu-1}(\xi)\cos(2Kx)].    
\end{equation}
\end{widetext}
From the general solutions in Eq.~\eqref{gen_sol}, the matrix elements of the time dependent perturbation $\cal{V}$, Eq. ~\eqref{armchair_selection_rule_general}, can be evaluated.

%

\end{document}